\documentclass{aa}  

\usepackage{graphicx}
\usepackage{txfonts}
\usepackage{threeparttable}
\usepackage{comment}
\newcommand{\lapprox }{{\lower0.8ex\hbox{$\buildrel <\over\sim$}}}
\newcommand{\gapprox }{{\lower0.8ex\hbox{$\buildrel >\over\sim$}}}
\def\amin{\ifmmode^{\prime}\else$^{\prime}$\fi}
\def\asec{\ifmmode^{\prime\prime}\else$^{\prime\prime}$\fi}

\usepackage[breaklinks, colorlinks, citecolor=blue, linkcolor=blue]{hyperref}

\begin{document}

   \title{Lithium, rotation and metallicity in the open cluster M35}

   
    \authorrunning{Cuenda-Muñoz et al.}
   \author{D.~Cuenda-Muñoz \inst{1}\fnmsep\thanks{dcuenda@cab.inta-csic.es},
          D.~Barrado \inst{1},
          M.~A.~Agüeros \inst{2,3},
          J.~L.~Curtis \inst{2},
          \and
          H. Bouy \inst{3,4}
          }

   \institute{Centro de Astrobiología (CAB), CSIC-INTA, ESAC Campus, Camino Bajo del Castillo s/n, 28692, Villanueva de la Cañada, Madrid, Spain\\
   \and
   Department of Astronomy, Columbia University, 550 West 120th Street, New York, NY 10027, USA\\
   \and
   Laboratoire d'astrophysique de Bordeaux, Univ.~Bordeaux, CNRS, B18N, All\'ee Geoffroy Saint-Hilaire, 33615 Pessac, France \\
   \and
   Institut Universitaire de France (IUF), Paris, France
             }

   \date{Received April 03, 2024; accepted May 04, 2024}

 
  \abstract
   {Lithium (Li) abundance is an age indicator for G, K, and M stellar types, as its abundance decreases over time for these spectral types. However, despite all of the observational efforts made over the past few decades, the role of rotation, stellar activity, and metallicity in the depletion of Li is still unclear.}
   {Our purpose is to investigate how Li depletion is affected by rotation and metallicity in G and K members of the roughly Pleiades-aged open cluster M35.}
  {We have collected an initial sample of 165 candidate members observed with the WIYN/Hydra spectrograph. In addition, we have taken advantage of three previous spectroscopic studies of Li in M35.
  As a result, we have  collected a final sample of 396 stars observed with the same instrument,  
  which we have classified as non-members, possible non-members, possible members, and probable members of the cluster. We have measured iron abundances, Li equivalent widths, and Li abundances for the 110 M35 members added to the existing sample by this study. Finally, rotation periods for cluster members have been obtained from the literature or derived from  Zwicky Transient Facility light curves.}
   {We have confirmed that fast G and K rotators are Li-rich in comparison with slow rotators of similar effective temperature. This trend, which is also seen in previous studies, is more evident when binaries are not taken into account. Furthermore, while we derived an average metallicity of [Fe/H] = $-$0.26$\pm$0.09 from our spectra, the distribution of Li in M35 is similar to those observed for the Pleiades and M34 open clusters, which have solar metallicity and slightly different ages. In addition, 
   we have shown that an empirical relationship proposed to remove the contribution of the Fe I line at 670.75 nm to the blended feature at 670.78 nm overestimates by 5--15 m$\AA$ the contribution of this iron line for M35 members.}
   {M35 fast G and K rotators  have depleted less Li than their slower counterparts. Furthermore, a 0.2$-$0.3 dex difference in metallicity appears to make little difference in the Li distributions of open clusters with ages between 100 and 250~Myr.}

   \keywords{stars: late-type --
                open clusters and associations : individual (M35, NGC 2168) --
                stars: rotation --
                stars: abundances
             }

   \maketitle
%

   \section{Introduction}

As lithium (Li) is gradually destroyed in the interiors of  solar-like F, G, and K main-sequence (MS) stars, the abundance of this element can be used as an age indicator for these stars. In addition, as the internal structure of the star has a strong influence on the depletion of Li, the study of the dependence of Li abundance  on spectral type and age provides a useful tool for improving stellar evolutionary models.

Li depletion is driven by convection, but additional factors play an important role in this process. For example, evidence of the connection between stellar rotation and Li depletion was found in the 1990s, when \cite{soderblom1993} revealed that the slow rotators in the Pleiades (125 Myr old) were Li-poor compared to the fast ones. As this discovery contradicted  predictions, many stellar associations have been observed in the decades since, and several different explanations have been proposed to account for this unexpected result.

\cite{siess1997} linked rotation and Li depletion through the mixing length parameter $\alpha$. A smaller $\alpha$ would imply a thinner convective envelope and, consequently, a reduction in the amount of Li that is destroyed. If $\alpha$ scales with the inverse of the angular velocity, rapid rotators would be richer in Li than slow rotators. On the other hand, \cite{Bou2008} proposed that the star-disk interaction, known as disk locking, that takes place during the early pre-main sequence (PMS) enhances differential rotation, preventing the star from spinning up. Consequently, a longer disk lifetime should result in a slower rotator that will reach the zero-age-main-sequence (ZAMS) with less Li due to this externally enforced rotational mixing. 

Multiple authors have studied the observed evolution of the Li--rotation connection from the PMS to the ZAMS. \cite{Bouvier2016} investigated this relationship in the 5-Myr-old, star-forming region NGC 2264, and demonstrated that the Li overabundance is detected in fast rotators even at this early stage of stellar evolution. \cite{BarradoPleiades2016} and \cite{BouvierPleiades2018}   confirmed the tight Li--rotation connection in the Pleiades, and \cite{Arancibia2020} also demonstrated its existence in the 125-Myr-old Psc-Eri stellar stream. These studies indicate that the Li overabundance increases during the PMS: Li abundance in fast rotators is 120$\%$ the Li abundance in slow rotators at 5 Myr, and it reaches 600$\%$ on the ZAMS, which is the evolutionary status of K dwarfs at the age of the Pleiades.

An additional consideration is the magnetic activity in these stars. \citet[][\citeyear{SomPis2015}]{SomPis2014} suggested that strong magnetic activity inflates the radius of young stars, reducing the temperature at the base of the convective envelope and, consequently, limiting PMS Li depletion. If fast rotators have stronger magnetic fields than slow ones, their Li abundance will consequently be higher. In this context, it is worth noting that while it is generally accepted that the observed Li spread for a given mass corresponds to real abundance differences, \citet[][and references therein]{BarradoPleiades2016} proposed that the Li overabundance in fast rotators is partially an observational effect from increased activity in there stars.

Finally, metallicity also influences Li depletion: stars with lower metallicity have thinner convective envelopes and, consequently, their Li depletion is less efficient. This anti-correlation between metallicity and Li depletion has recently been confirmed for solar analogs by \cite{Martos2023}, strengthening the models proposed by \cite{Castro2009} and \cite{Durmont2021}.

To completely understand the role that age, activity, rotation, and metallicity play in Li depletion, it is crucial to have a reliable and extensive set of observations of MS FGK stars. As the members of open clusters share the same age and metallicity, they constitute the perfect targets for these observations. Our focus is on the open cluster M35 (NGC 2168), a rich \citep{Ba2001b, Bo15}, relatively distant \citep[885~pc;][]{Je21} cluster that is often thought of as a much richer analog of the Pleiades.
 
\cite{Je21} have investigated the Li--rotation connection for G and K members of M35, finding a strong correlation between the equivalent width of the Li line, EW(Li), rotation, and radius inflation.
However, the position of the cluster in the age--metallicity plane is still ambiguous: while \cite{Je21} quote an age of 140$\pm$15 Myr, \cite{Abdelaziz2022} derived an age of 126 Myr for M35, closer to that of the Pleiades. More significantly, only a few authors have provided values for M35's  metallicity: \cite{Ba01a} and \cite{steinhauer2004} found sub-solar values ([Fe/H]~=~$-$0.21$\pm$0.10 and $-$0.143$\pm$0.014, respectively), while, based on their measurements of the EW of the Ca I line for stars in their sample, \citet{At18} assumed a solar metallicity for the cluster. 

In this paper, we present our analysis of unpublished spectra and new light curves, with which we have expanded the sample of M35 members with EW(Li) and rotation period measurements presented in \cite{Je21}. In addition, we  derive the metallicity of M35 by analyzing the spectra collected for a sample of single GK dwarfs with very high probability of being cluster members. Finally, we have compare our results for M35 to those for the Pleiades and for M34, another open cluster with an age quite similar to that of M35.

In Section~\ref{sec:spec}, we present our spectroscopic sample, obtained with multiple nights of observations using the Hydra spectrograph on the 3.5-m telescope, WIYN Observatory, Kitt Peak, USA.\footnote{The WIYN Observatory is a joint facility of the NSF's National Optical-Infrared Astronomy Research Laboratory, Indiana University, the University of Wisconsin-Madison, Pennsylvania State University, the University of Missouri, the University of California-Irvine, and Purdue University. The astronomical community is honored to have the opportunity to conduct astronomical research on Iolkam Du’ag (Kitt Peak) in Arizona. We recognize and acknowledge the very significant cultural role and reverence that this site has to the Tohono O'odham Nation.} We also describe other WIYN/Hydra M35 campaigns whose results were incorporated into this study. In Section~\ref{sec:mem}, we describe our efforts to finalize a membership catalog for M35, and to distinguish between candidate single and candidate binary members of the cluster. In Section~\ref{sec:rot}, we discuss our sample of rotational period measurements for the members of M35. This includes new rotation periods obtained from our analysis of Zwicky Transient Facility \citep[ZTF;][]{masci2019} light curves, as well as periods from the literature. Section \ref{sec:spec_analysis} explains the method employed to obtain effective temperatures and luminosities for the members of M35, and describes the analysis carried out to derive the metallicity of this cluster and provide new Li measurements. Finally, the results obtained are presented and discussed in Section \ref{sec:results}, and we provide our conclusions in Section \ref{sec:conclusions}.

   \section{Assembling our spectroscopic  sample}\label{sec:spec}

Our sample is made up of M35 candidate members observed with the Hydra multifiber spectrograph. Hydra has more than 80 fibers with which one can cover a field 60 arcminutes in diameter. Since half of M35's members are contained within a diameter of less than 40 arcminutes \citep{Cantat2020}, Hydra is able to observe most of the cluster within a single observation.
The light collected by each fiber is transmitted to the bench spectrograph and scattered by the selected grating. Two different science cables are available for each configuration of fibers, the red one covering 400--1100~nm and the blue covering 300--800 nm.

\subsection{Previously unpublished Hydra spectra}\label{sec:new_sp}

Over the course of four observational campaigns between 1998 January and 2001 March, more than 300 M35 candidate members were observed with Hydra. Here we focus on the two campaigns with unpublished data. From 1999 December 12--15, eight science images were taken using the blue Hydra fiber cable, each with the same configuration of 17 sky fibers and 76 science fibers. From 2001 February 22--23, five science images were taken using the same fiber cable, but the configuration was slightly different: there were five sky fibers and 89 science fibers. None of the sources observed in 1999 was observed again in 2001 February. Table \ref{table:1} list the dates of the observations, the central coordinates, and the exposure times for each image.

\begin{table}
\begin{threeparttable}
\caption{Science exposures taken during the December 1999 and February 2001 WIYN/Hydra campaigns. Date, central coordinates, and exposure time are shown for each exposure.}            
\label{table:1}      
\centering      
\begin{tabular}{lcccc}   
\hline\hline        
Image & $\alpha$(2000) & $\delta$(2000) & Date & Exposure \\ 
 & hh:mm:ss & dd:mm:ss &  & Time (s) \\
\hline                    
   N2017 & 06:09:06.0 & 24:20:42 & 1999 Dec 12 & 4355 \\
   N2018 & 06:09:06.1 & 24:20:42 & 1999 Dec 12 & 7200 \\
   N2019 & 06:09:06.1 & 24:20:42 & 1999 Dec 12 & 7200 \\
   N2020 & 06:09:06.1 & 24:20:42 & 1999 Dec 12 & 4920 \\
   N3033 & 06:09:06.0 & 24:20:42 & 1999 Dec 13 & 7200 \\
   N3034 & 06:09:06.0 & 24:20:42 & 1999 Dec 13 & 4500 \\
   N4049$^*$ & 06:09:06.0 & 24:20:42 & 1999 Dec 14 & 2725 \\
   N5036 & 06:09:06.0 & 24:20:42 & 1999 Dec 15 & 6300 \\
   N1033 & 06:09:04.4 & 24:18:48 & 2001 Feb 22 & 7200 \\  
   N1034 & 06:09:04.4 & 24:18:48 & 2001 Feb 22 & 7200 \\
   N1036 & 06:09:04.4 & 24:18:48 & 2001 Feb 22 & 7800 \\
   N2099 & 06:09:04.4 & 24:18:48 & 2001 Feb 23 & 7200 \\
   N2100 & 06:09:04.4 & 24:18:48 & 2001 Feb 23 & 9000 \\
\hline       
\end{tabular}
\begin{tablenotes}
\small
\item $^*$We used N4049 to derive RVs, but not to measure Li EWs.
\end{tablenotes}
\end{threeparttable}
\end{table}

We reduced the 165 stellar spectra obtained as follows. For homogeneity with previous studies with Hydra (section \ref{sec:Ba01_At18}), we used standard IRAF\footnote{IRAF is distributed by the National Optical Astronomy Observatories, which are operated by the Association of Universities for Research in Astronomy, Inc., under cooperative agreement with the National Science Foundation.} tools to obtain the smoothed and continuum-normalized spectrum corresponding to each source in each of the images. First, we smoothed the spectra taken during each exposure using the boxcar task included in the images.imfilter package. Second, we removed the continuum by fitting a cubic spline function to each spectrum taking advantage of the continuum task included in noao.onedspec. Finally, we employed the scopy task in imred.hydra to extract the individual (smoothed and continuum normalized) spectrum for each observed source. Each of the resulting \textit{R} $\approx$ 20,000 spectra covers the wavelength interval between 640 and 690 nm. The Li doublet at 670.78 nm is included in this range as well as a number of iron lines.

\subsection{Previously published spectra from these campaigns }\label{sec:Ba01_At18}
 
We also included in our sample the G and K dwarfs observed in the two other observing campaigns from that same time period and analysed in previous publications \citep{Ba01a,At18}. Although these authors address the Li--rotation connection, they rely on rotation velocities, which are affected by the uncertainty in the rotation inclination angle, in contrast to the analysis provided in this work, which relies on measurements of rotation periods.

\cite{Ba01a} observed 76 candidate members of M35 selected from the photometric survey of the cluster of \cite{Ba2001b}. \cite{At18} acquired spectra of 85 stars, 77 of which had preliminary radial velocities (RVs) consistent with membership; the other eight were selected based on their positions in a color--magnitude diagram (CMD). There is no overlap between the objects analysed in the aforementioned studies, but a few of them were also observed in  1999 December and 2001 February. Fifteen of the spectra presented in this work are for stars also observed by \cite{Ba01a}, and five are also included in \cite{At18}.

   \subsection{Other Hydra surveys of M35} \label{sec:Jeff21}

\cite{Je21} employed Hydra to observe 342 candidate members of M35 selected from the catalogs of \cite{Bo15}, \cite{Meibom2009}, and \cite{Libralato2016}. \cite{Je21} rejected 100 of these stars because they considered them non-members of the cluster, found no rotation periods for them, or these objects exhibited a poor fit when analysed with the Virtual Observatory Spectral Energy Distribution (SED) Analyser \citep[VOSA;][]{Bayo2008}. Because we developed our own set of membership criteria, and have a different set of rotation period measurements for our analysis, we have incorporated these 100 discarded stars into our sample (ten were observed as part of our program).

On the other hand, the 242 stars that were analyzed by \cite{Je21} have not been included in our sample for re-analysis. Instead, we incorporated the \cite{Je21} results for these stars into our analysis in section \ref{sec:results}. It is worth mentioning that 115 of these stars are included in \cite{Ba01a}, \cite{At18}, or the Hydra campaigns presented in this work.

   \subsection{The complete sample}

As shown in Table \ref{table:sample_origin}, our sample includes 76 stars taken from \cite{Ba01a}, 85 taken from \cite{At18}, 100 taken from \cite{Je21}, and 165 stars added by this study. As a result, we obtained a final sample of 396 unique G and K candidate members of M35 for which we could obtain or have Li measurements. The complete sample is presented in Table \ref{table:sample}.

\begin{table}
\caption{Summary of our complete sample in terms of the origin of the data.}      
\label{table:sample_origin}      
\centering          
\begin{tabular}{l c c}     
\hline\hline        
Campaign & Number of & Authors \\
 & Sources & \\
\hline                    
   1998 January & 76 & \cite{Ba01a} \\  
   1999 December & 76 & This Work \\ 
   2001 February & 89 & This Work \\
   2001 March & 85 & \cite{At18} \\
   2017 November & 100 & \cite{Je21} \\ 
\hline                  
\end{tabular}
\end{table}

\begin{table*}
\caption{Cross-identification numbers, coordinates, and observation campaigns for the complete sample presented in this work. Our sample is composed of 396 stars. Columns $\alpha$ and $\delta$ indicate the coordinates employed for the pointing. For the stars that have a counterpart in \citet{At18} and/or \citet{Je21}, we have also included the corresponding IDs in these papers. This table is only partially
presented. Its complete version is available electronically.}             
\label{table:sample}      
\centering          
\begin{tabular}{c c c c c c c c}     
\hline\hline        
ID \tablefootmark{a} & $\alpha$(2000) & $\delta$(2000) & $\alpha$(2000) & $\delta$(2000) & Campaign & AT18\tablefootmark{b} ID & Je21\tablefootmark{c} ID \\
 & hh:mm:ss & dd:mm:ss & deg & deg &  &  &  \\
\hline                    
   5076 & 06:08:52.9 & +24:17:20 & 92.22042 & 24.28889 & 2001 Mar & 16010 & -- \\
   5081 & 06:09:26.200 & +24:29:03.0 & 92.35917 & 24.48417 & 1998 Jan & -- & -- \\ 
   7117 & -- & -- & 92.22672 & 24.05225 & 2017 Nov & -- & J06085441+2403081 \\
   7036 & 06:09:00.7 & +24:35:53 & 92.25292 & 24.59806 & 2001 Feb & -- & -- \\
   5373 & 06:08:49.270 & +24:15:33.30 & 92.20529 & 24.25925 & 1999 Dec & -- & --  \\
   ... & ... & ... & ... & ... & ... \\ 
\hline                  
\end{tabular}
\tablefoot{
As the 20 objects in common between the campaigns analysed in this work and the ones presented in \cite{Ba01a} and \cite{At18} are shown as different rows with the same identification number, the table includes 416 rows.
\tablefoottext{a}{Object IDs between 5000 and 7000 are taken from \cite{Ba2001b}. The stars not found in that catalog have been labelled with numbers between 7000 and 8000.}
\tablefoottext{b}{AT18 is \citet{At18}.}
\tablefoottext{c}{Je21 is \citet{Je21}.}
}
\end{table*}

	\section{Revisiting the membership and multiplicity of stars in our sample}\label{sec:mem}

\subsection{An initial membership selection based on \cite{Bo15} and our own proper-motion analysis}

Using TOPCAT \citep{taylor2005}, we cross-matched our 396 stars with the \textit{Gaia} Data Release 3 \citep[DR3;][]{gaiaDR3}, obtaining astrometry and photometry for 395 candidate members (see the CMD in the left panel, Figure \ref{CMD_Clusterix}). We also cross-matched our catalog with that of \cite{Bo15}, as these authors provided an extensive membership analysis for M35. We found counterpart for 373 stars in that work.

We then used Clusterix 2.0, a Spanish Virtual Observatory tool \citep{2020MNRAS.492.5811B}, to assign membership probabilities to our candidates based on their \textit{Gaia} Early Data Release 3 \citep[EDR3;][]{gaiaEDR3} proper motions.\footnote{Clusterix 2.0 does not yet include DR3 proper motions.} After selecting a region of the sky including M35 and a region free of cluster members, Clusterix 2.0 uses a non-parametric method to distinguish the proper-motion distribution of the cluster members and that of the field stars. In this process, the number of cluster members, \textit{N$_c$}, is estimated, and the membership probability of each star in the selected regions is computed. 

Clusterix 2.0 returns a file that includes, for each star considered, the \textit{Gaia} EDR3 photometry and astrometry and a membership probability. In addition, the \textit{N$_c$} stars with the highest membership probabilities are flagged as members of the cluster. We removed from this preliminary list of candidate members around half of them due to the large uncertainties in the photometry and/or astrometry measured for those stars, or because their parallaxes and/or proper motions were very different from those of M35 \citep{2018A&A...616A..10G}.

We then combined the membership probabilities of \cite{Bo15} with those derived by Clusterix 2.0 to classify each star as a non-member, a possible non-member, a possible member, or a probable member. 

We flagged as non-members stars whose membership probabilities were below 50$\%$ in \citet{Bo15} and that were also rejected as members after running Clusterix 2.0. 
In addition, visual inspection of the cluster CMDs allowed us to identify 16 stars as photometric non-members. The stars that had a counterpart in either \citet{Bo15} or Clusterix 2.0, but not in both, and whose membership probabilities did not meet our requirements, were classified as possible non-members, as were stars that did not have a counterpart in either of them. 

We classified as possible members the stars that were considered members by \citet{Bo15}, that is, their membership probabilities were $\geq$50$\%$, or after running Clusterix 2.0, but not by both. Finally, we classified as probable members the stars that met both requirements.

In this manner, we identified 101 stars in our original sample of 396 as non-members (26\%), 44 stars as possible non-members (11\%), 129 stars as possible members (32\%), and 122 as probable members (31\%) of M35. 
In our catalog, presented in Table~\ref{table:membership_binaries}, we use the labels \textit{NM}, \textit{CMD NM}, \textit{NM?}, \textit{Poss}, and \textit{Prob} to indicate these membership categories.

   \subsection{Measuring radial velocities and distinguishing between single and binary stars in our sample} \label{radial_velocities}

\subsubsection{Measuring radial velocities from our Hydra data}

We employed iSpec, described in \cite{2014A&A...569A.111B}, to measure radial velocities (RVs) for the M35 candidate members observed during the 1999 December and 2001 February campaigns. First, each spectrum was transformed to the solar barycentric reference frame. We then derived the RV by cross-correlating the spectrum with a solar mask covering the wavelength range 372--926 nm. We obtained eight measurements of the RV for the M35 candidates observed in 1999 and five measurements for those observed in 2001.

We used these RVs to distinguish between spectroscopic binaries and single stars by employing an approach similar to that of \citet{Le15}. The standard deviation in RV ({\it e}) for each star was compared with the maximum and the minimum single-measurement precision in RV calculated by iSpec for that star ({\it max(i)} and {\it min(i)}, respectively). 

The stars whose ratios {\it e/max(i)} and {\it e/min(i)} were both $<$4 were considered single, while those whose standard deviation in RV was clearly above the single-measurement precision derived for any of their spectra were classified as SBs. The M35 candidates with {\it e/max(i)} $<$ 4 but {\it e/min(i)} $\geq$ 4 were classified as having an unknown multiplicity. 

   \begin{figure*}[!th]
   \centering
   \includegraphics[trim=0.2cm 0.2cm 0.2cm 0.75cm, clip=true, width=\columnwidth]  {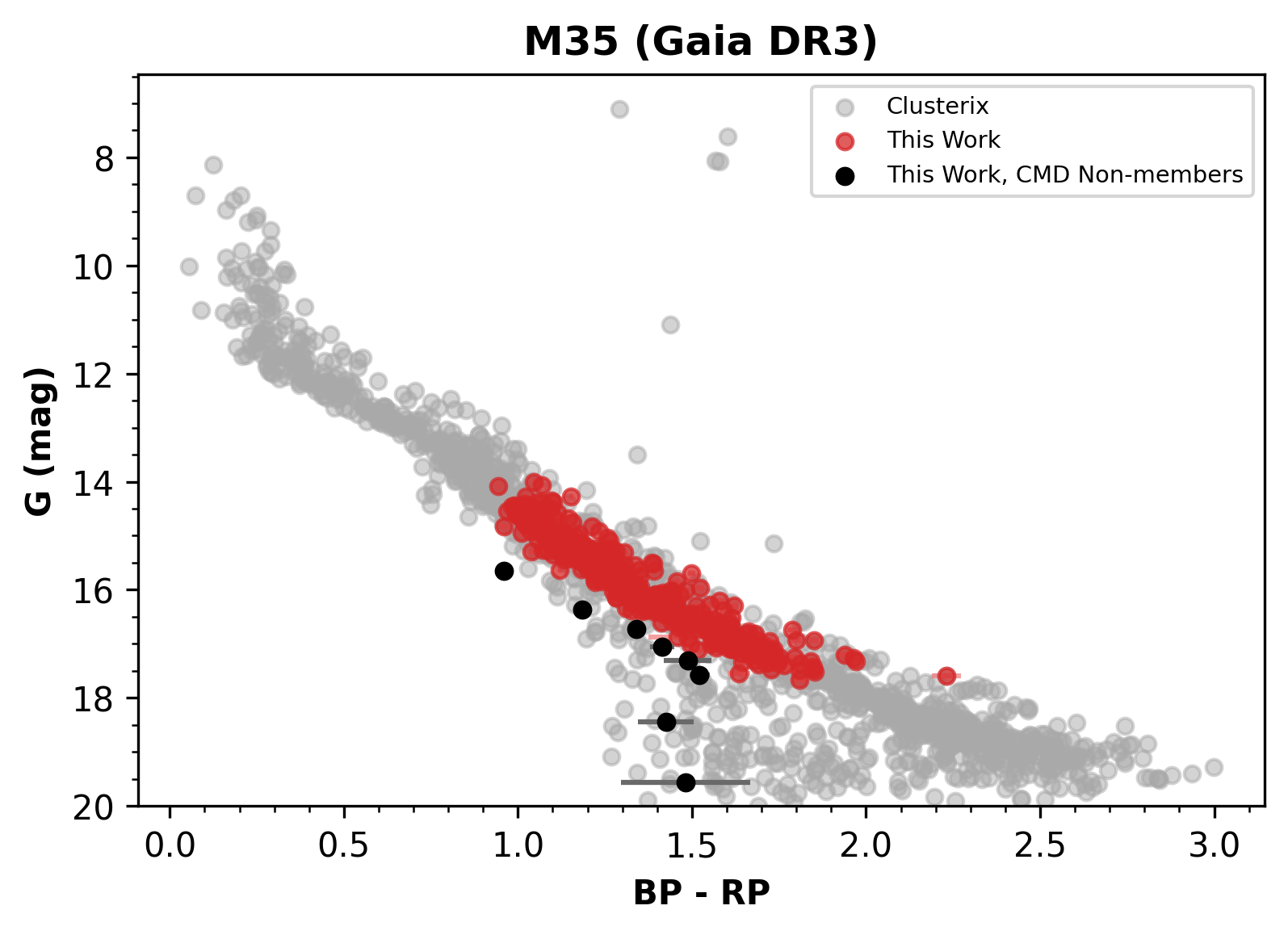}
   \includegraphics[trim=0.2cm 0.2cm 0.2cm 0.65cm, clip=true, width=\columnwidth, height=6.15cm]{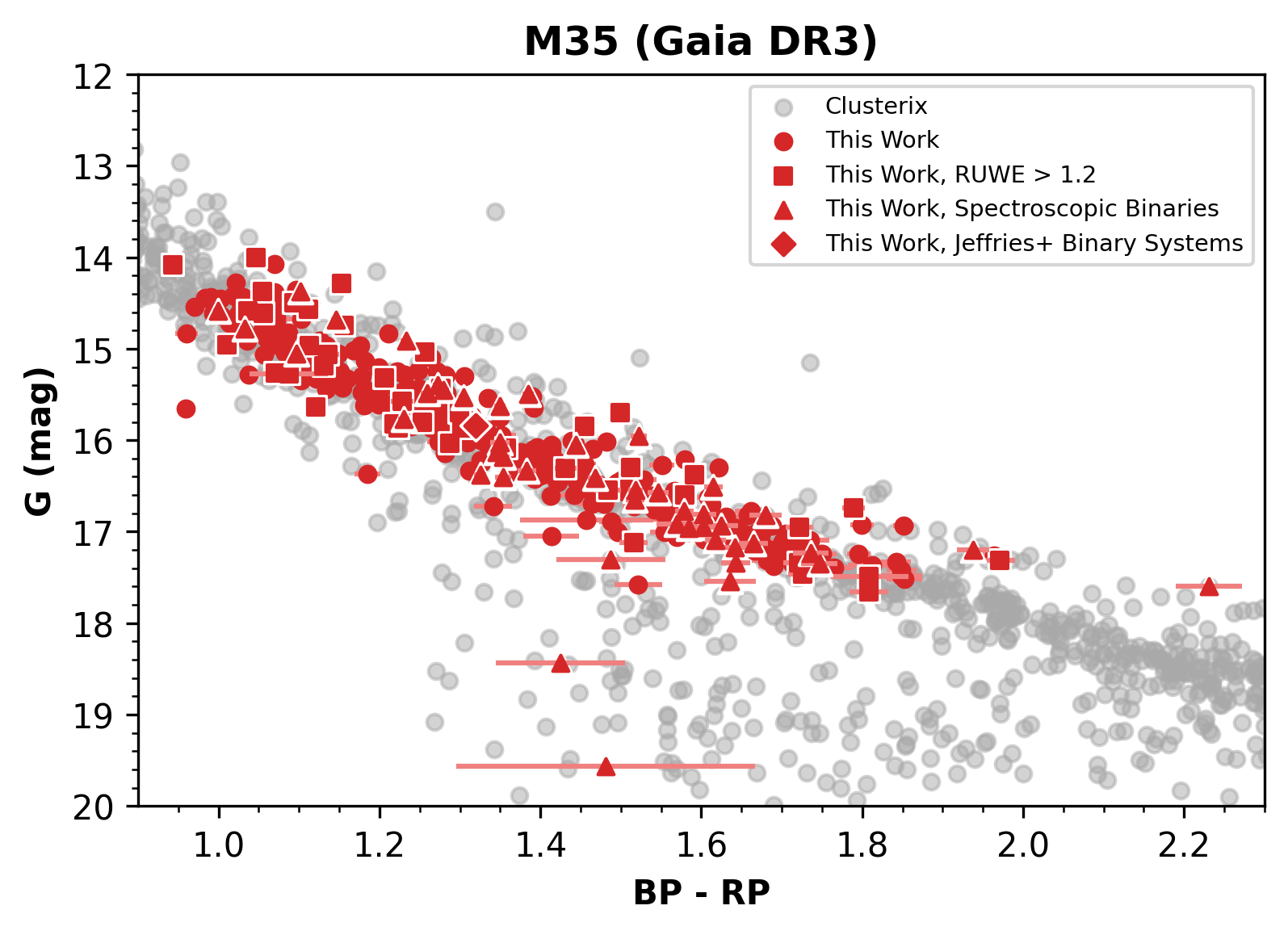}
   \caption{\textit{Gaia} DR3 CMD for M35. \textit{Left---}The grey points represent stars in the vicinity of M35 that Clusterix 2.0 classified as cluster members. We have carried out a subsequent analysis removing some stars whose parallaxes and/or proper motions were very different from those of M35 \citep{2018A&A...616A..10G}. Candidate members of the cluster in our sample are shown in red, while the black circles are stars whose photometry suggests they are not members of M35. We used optical and infrared photometry published in \citet{Bo15} to classify eight other photometric outliers as non-members. {\it Right---}A zoomed-in version of the CMD highlighting the multiple systems in our sample. The squares are potential wide binaries identified by \textit{Gaia}, triangles are spectroscopic binaries, and diamonds are binary systems found by \citet{Je21}. Note that neither CMD has been corrected for  reddening.}
      \label{CMD_Clusterix}%
    \end{figure*}

In addition, the average RV of each single star was calculated to distinguish between single members and single non-members of the cluster. We considered single members those stars whose average RVs fall within the interval [$V_{rad} - 3\sigma$, $V_{rad} + 3\sigma$], where $V_{rad}$ and $\sigma$ were taken from \cite{2018A&A...616A..10G}: $V_{rad} = -7.70$ km s$^{-1}$; $\sigma = 0.27$ km s$^{-1}$. 

We added a ``$-$'' symbol to the membership labels employed for the non-members and possible non-members in our sample that were also considered single non-members based on their RVs. Conversely, we added a ``$+$'' to the membership labels for the possible and probable members whose RVs indicate that they are single members of M35.

\subsubsection{Using Gaia data and the literature to distinguish between single and binary stars}

\textit{Gaia} DR3 and previous studies of M35 provided other avenues for identifying single and binary members of the cluster, particularly for those stars for which we did not obtain our own Hydra spectra. The \textit{Gaia} Renormalised Unit Weight Error (RUWE) is expected to be $\lapprox$1.0 for sources when the single-star model gives a good fit to the astrometric observations, and previous studies have found that larger RUWE values imply that the objects are unresolved binaries in wide orbits \citep[e.g.,][]{Deacon2020, Ziegler2020, Kervella2022}. Accordingly, we flagged as wide binaries the 41 stars in our sample whose RUWE $>$ 1.2. 

We also took advantage of the membership classification published in \cite{Ba01a} for the 76 M35 candidate members  observed by these authors. After analyzing the RVs they measured, these authors split their sample into single members (39 stars), spectroscopic binary members (13), and non-members (24). We employed these labels for the part of the sample taken from that work but we kept the classification described in the previous section for the 15 stars observed in 1999 December and 2001 February that are included in \citet{Ba01a}. Six of these 15 dwarfs are members of M35.

As a further verification of our membership scheme, we cross-matched our complete sample with the catalog of \citet{Le15}, finding counterparts for 174 of the 396 stars. Reassuringly, the 28 stars that we classify as single members that have counterparts in \citet{Le15} are also M35 members in that work. We turned to the binary / single star classification provided by \citet{Le15} only for the part of the sample taken from \citet{At18} and the part recovered from \citet{Je21}.  

At the end of this process, we had the sample of 41 wide binaries determined from their RUWE values, and another 46 stars identified as spectroscopic binaries thanks to the RV analysis discussed. One more binary system in our sample was identified from checking the list of probable binary systems defined by \citet{Je21}. In total, we therefore found that 88 of the 396 candidates members that composed our sample could be considered binaries. 
These 88 binaries are represented by squares, triangles, and diamonds in the CMD shown in the right panel of Figure~\ref{CMD_Clusterix}, and are indicated with binary flag = 1 in Table ~\ref{table:membership_binaries}.

   \subsection{Summary}

As mentioned above, we removed 16 candidates from our sample after visual inspection of their position in several CMDs. 
Eighty-five other stars  were discarded because the membership probabilities derived by \cite{Bo15} and the analysis carried out with Clusterix 2.0 both labeled them as non-members. 
Reassuringly, 28 of these 85 candidates (labeled with \textit{NM-} in Table~\ref{table:membership_binaries}) have RVs incompatible with M35 membership. Similarly, 17 of the 44 stars classified as possible non-members have RVs incompatible with cluster membership. These 17 dwarfs are labeled with \textit{NM?-} in Table~\ref{table:membership_binaries}.

On the other hand, we identified 129 dwarfs as possible members.  
The RVs confirmed 61 of these candidates as cluster members (labeled with \textit{Poss+} in Table~\ref{table:membership_binaries}). In addition, our sample includes 122 M35 probable members, 
of which 68 (labeled with \textit{Prob+} in Table~\ref{table:membership_binaries}) are single stars with RVs consistent with that of the cluster. Hereafter, we refer to the combined list of 251 possible and probable members we obtained as members of M35.

\begin{table}
\caption{Membership class and binary status for the complete sample presented in this work. Our sample is composed of 396 stars. This table is only partially presented. Its complete version is available electronically.}
\label{table:membership_binaries}      
\centering          
\begin{tabular}{c c c c c c}     
\hline\hline        
ID & RV & RV & Final Memb. & Binary \\ 
 &  Class \tablefootmark{a} & Source \tablefootmark{b} & Class & Flag \tablefootmark{c} \\
\hline                    
   5076 & SM & L15 & Poss+ & 0 \\  
   5081 & SB1 & B01 & NM & 1 \\ 
   7117 & BM & L15 & Prob & 1 \\
   7036 & SM & This Work & Prob+ & 0 \\
   5373 & U & This Work & CMD NM & 0 \\
   ... & ... & ... & ... & ... \\ 
\hline                  
\end{tabular}
\tablefoot{
\tablefoottext{a}{\textit{MEM} and \textit{SM} account for single members; \textit{NM} and \textit{SN} account for single non-members; \textit{SB}, \textit{BLM}, \textit{BLN}, \textit{BM}, \textit{BN}, \textit{BU}, \textit{SB1}, and \textit{SB2} account for spectroscopic binaries; and \textit{U} accounts for unknown multiplicity.}
\tablefoottext{b}{B01 is \cite{Ba01a} and L15 is \citet{Le15}.}
\tablefoottext{c}{1 = likely binary, 0 = single star.}
}
\end{table}

	\section{Assembling our sample of rotation periods}\label{sec:rot}

We cross-matched our sample of 251 cluster members with the photometric surveys of \cite{Meibom2009}, \cite{N15}, \cite{Libralato2016}, and  \cite{SF20}. 
In addition, new rotation periods were derived by analyzing the light curves obtained from Zwicky Transient Facility \citep[ZTF;][]{masci2019} data. The main properties of these photometric surveys are shown in Table \ref{table:phot_surveys}.

\begin{table*}
\caption{Field of view, instrument characteristics, timespan, and cadence for the photometric surveys from which we obtained rotation periods for our sample.}             
\label{table:phot_surveys}      
\centering          
\begin{tabular}{l c c c c c}     
\hline\hline 
Authors		 	& Telescope 	& Field of View 		& Plate scale 		& Timespan 				& Cadence  \\ 
 			&  		&  				& (arcsec/pixel)	&  					& \\
\hline                    
\citet{Meibom2009}	& WIYN 0.9 m 	& $\approx$20.5\amin x20.5\amin & 0.62			&  16 / 143\tablefootmark{a} d  	& 1 hr / 1 d \\ 
    			& (Kitt Peak Observatory) &  			& 			&					& \\   
\citet{N15} 		& Asiago 67/92 cm Schmidt & $\approx$58\amin x38\amin & 0.86 		& $\geq$10 d 			& 3 min  \\
    			& (OAPD) 		& 			&			&					& \\
\citet{Libralato2016} 	& Kepler (K2)		& 115$^{\circ^2}$ 	& 3.98			& $\approx$31 d 			& 1 min / 29 min\tablefootmark{b}  \\
			&  			&  			&			& 					& \\
\citet{SF20}		& Kepler (K2)		& 115$^{\circ^2}$	& 3.98			& $\approx$31 d 			& 29 min \\
			&  			&  			&			& 					& \\
This Work 		& Samuel Oschin Telescope & 47$^{\circ^2}$	& 1.00 			& several yrs 			& 1 - 2.5 d \\
			& (Palomar Observatory; ZTF) & 			&			&					& \\
\hline  
\end{tabular}
\tablefoot{
\tablefoottext{a}{High-frequency (once per hr for 5--6 hrs per night) time-series photometric observations were taken over 16 nights and complemented with low-frequency (once per night) observations taken over 143 nights.}
\tablefoottext{b}{\textit{Kepler} exposures are combined on board to create short-cadence timeseries (nine co-added exposures) or long-cadence timeseries (270 co-added exposures). See \cite{2010ApJ...713L..79K} and references therein for a detailed description of the \textit{Kepler} design.}
}
\end{table*}

   \subsection{Rotation periods from the literature}\label{Prot_Lit}

Our primary source of literature periods is the survey of \citet{Meibom2009}, who mixed high-frequency observations over the course of about two weeks with low-frequency observations over nearly five months to measure periods for both fast and slow rotators in M35. By matching our catalog with that of \citet{Meibom2009}, we obtained periods for 109 stars in our sample. A similar choice was made by \citet{Je21}, who chose the periods published in \citet{Meibom2009} instead of the ones derived by \citet{Libralato2016} when both were available. 

For those stars lacking periods in \citet{Meibom2009}, we turned to the ones derived from K2 light curves. From \citet{SF20}, we selected periods for the 31 stars classified as rotating with no blending. In addition, we adopted the period published for one star classified as indeterminate variable.

We have taken advantage of the periods not flagged as possible blends in \citet[][flags 0 and 30]{Libralato2016} to obtain rotation periods for 39 stars not included in \citet{SF20}. In addition, we recovered one period from \citet{Libralato2016} for a star classified in \citet{SF20} as an eclipsing binary.

Finally, we obtained rotation periods for six stars from \citet{N15}. Two have no counterparts in \citet{Meibom2009}, \citet{Libralato2016}, \citet{SF20}; one has an invalid period (\textit{Prot} = $-$99.0) in \citet{Libralato2016}; and three are considered  blends in \citet{SF20}. As \citet{N15} analysed light curves taken by an instrument whose plate scale is much smaller than \textit{Kepler}'s, we considered that these light curves were unlikely to be affected by blending and that these periods are reliable.

\subsection{New rotational periods from ZTF light curves}

ZTF has observed M35 regularly since 2018 March 25. Following \citet{curtis2020}, we extracted differential photometry from the archival ZTF imaging using nearby reference stars in the field. Although the resulting light curves are much sparser and more irregularly sampled than those from dedicated photometric surveys, they can still yield reliable period measurements. Our goal in analyzing these light curves was twofold: ZTF-derived periods can provide confirmation of periods in the literature, and ZTF light curves generally cover much longer timespans than those of other photometric surveys, and can therefore be used to measure periods for slower rotators in M35.

We inspected the light curves for each available season, computed Lomb--Scargle periodograms, and then recorded the period if it was significant and produced a convincing phase-folded light curve. In this manner, we measured periods for nine stars lacking periods in the literature, for one star classified as rotating with blending in \citet{SF20}, and for 37 stars with valid periods in the aforementioned surveys. For stars with periods in the literature, the average agreement between our measurements and the existing values is $<$1\%, and the differences are $\leq$10\%. We have chosen the periods derived from ZTF light curves for these 47 members of M35.

   \subsection{Final sample of M35 members with Li measurements and rotation periods}\label{All_Prot}

Combining the periods we derived and those collected from the literature, we obtained rotation periods for 197 of the 251 members of M35 observed with Hydra. Table~\ref{table:Prot} provides the periods for these 197 stars.

\begin{table*}
\caption{Rotation periods taken from the literature or derived from ZTF light curves for 197 M35 members. M09 is \citet{Meibom2009}, SF20 is \citet{SF20}, N15 is \citet{N15}, and L16 is \citet{Libralato2016}. This table is only partially presented. Its complete version is available electronically.}             
\label{table:Prot}      
\centering          
\begin{tabular}{c c c c c c c c c c c c}     
\hline\hline        
ID & Prot (d) & Prot (d) & Class \tablefootmark{a} & Subclass \tablefootmark{b} & Blend \tablefootmark{c} & Prot (d) & Flag \tablefootmark{d} & Prot (d) & Type \tablefootmark{e} & Prot (d) & Final \\
 & M09 & SF20 & SF20 & SF20 & SF20 & L16 & L16 & N15 & N15 & ZTF & Prot (d) \\ 
\hline                    
   5483 & -- & -- & -- & -- & -- & 0.48 & 1 & 0.48 & Rot & 0.48 & 0.48 \\
   7118 & 1.10 & 1.11 & Rotating & RotVar & 0 & 1.11 & 1 & 1.10 & Rot & 1.11 & 1.11 \\
   5356 & -- & -- & -- & -- & -- & -- & -- & -- & -- & 7.01 & 7.01 \\
   5459 & 7.30 & -- & -- & -- & -- & 7.19 & 1 & -- & -- & -- & 7.30 \\
   7076 & -- & 2.44 & Rotating & RotVar & 0 & -- & -- & -- & -- & -- & 2.44 \\
   ... & ... & ... & ... & ... & ... & ... & ... & ... & ... & ... & ... \\
\hline                  
\end{tabular}
\tablefoot{
\tablefoottext{a}{\textit{Rotating} accounts for rotational variables, \textit{Pulsating} for pulsating variables, \textit{EB} for eclipsing binaries, and \textit{Misc} for indeterminate variables.} 
\tablefoottext{b}{\textit{RotVar} accounts for rotational variables of indeterminate type, \textit{GDor} for $\gamma$ Doradus pulsators, \textit{EB} for eclipsing binaries, and \textit{Misc} for indeterminate variables.}
\tablefoottext{c}{0 = identified primary variable, 1 = ambiguous blend.}
\tablefoottext{d}{0 = high probability that it is a blend; 1 = candidate variable; 2 = difficult to classify; 30 = difficult to classify star that, by comparison with the literature, could be a possible blend; 31 = difficult to classify star for which a correspondence in the literature was found.}
\tablefoottext{e}{\textit{Rot} are rotating stars, \textit{EB}  eclipsing binaries, \textit{Long-Period}  long-period variables, and \textit{{delta} Sct} $\delta$ Scuti stars.}
}
\end{table*}

	\section{Analysis}\label{sec:spec_analysis}

   \subsection{Deriving effective temperatures and luminosities}\label{VOSAfit}

We generated the SED of each star in our sample with VOSA \citep[version 7.5;][]{Bayo2008} using the photometry in Table~\ref{table:photometry}, which was taken from \citet{Bo15} and \cite{Ba2001b}. We also used VOSA to cross-match our sample with a number of catalogs, obtaining photometry in the PAN-STARRS/PS1 filters, Misc/APASS filters, WISE filters, INT/IPHAS filters, and Palomar/ZTF filters. Following an approach similar to that taken by \citet{Je21}, we assumed a distance of 885$\pm$80 pc for all the stars and a visual extinction in their direction of \textit{$A_{V}$} = 0.62. 

\begin{figure}
   \resizebox{\hsize}{!}{\includegraphics{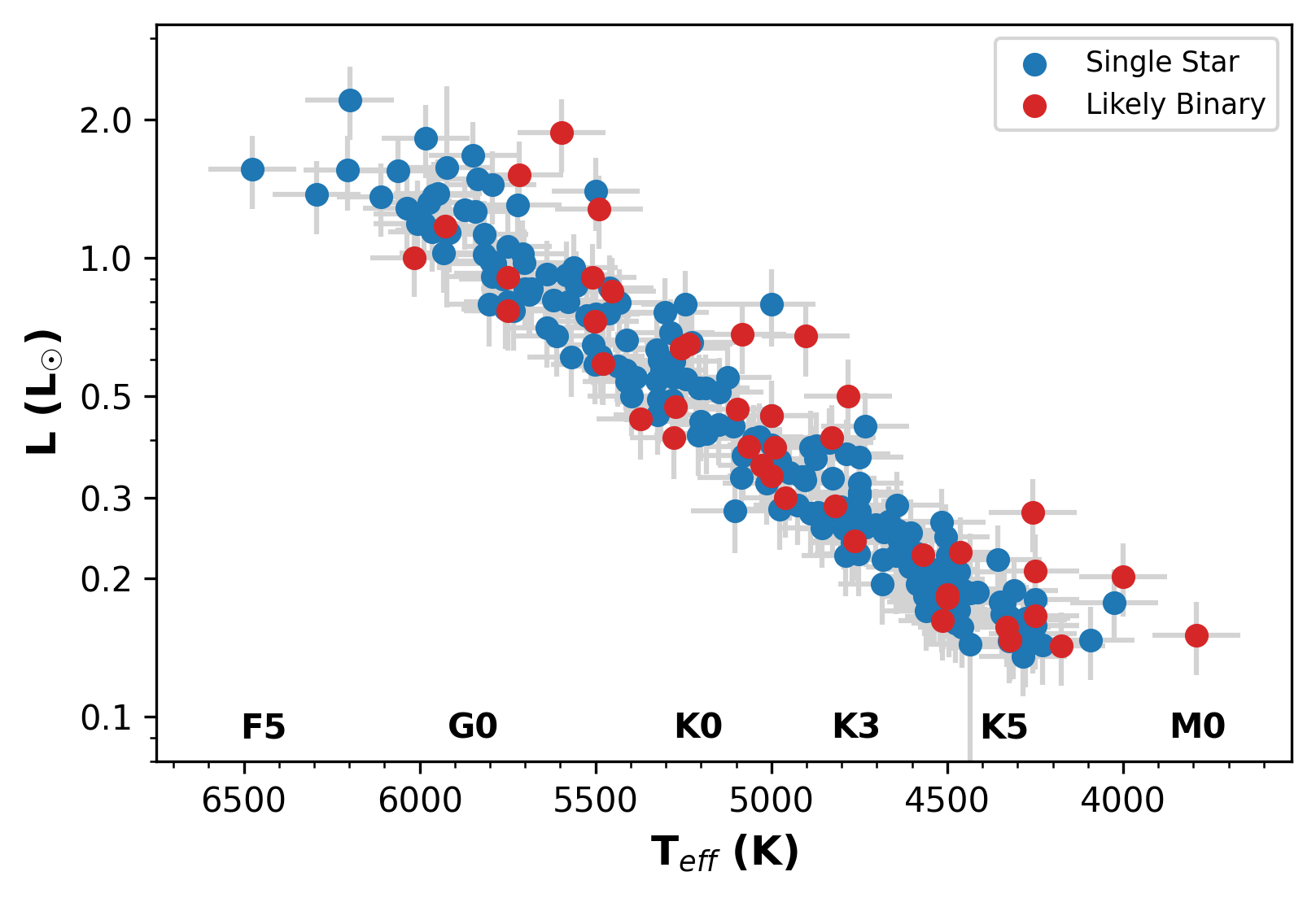}}
   
   \caption{Positions in the HRD for the 250  members of M35 in our sample whose effective temperatures and bolometric luminosities were determined by SED fitting with VOSA. Red circles are binary systems, while blue circles are single stars.}
              \label{HRdiagram_VOSA}%
    \end{figure}

The dereddened SEDs obtained in this manner were compared with Kurucz model atmospheres \citep[]{CasKur1997, CasKur2003}, assuming log \textit{g} = 4.5 and a solar metallicity, to determine the luminosity and effective temperature that best fit each SED using a $\chi$-squared minimisation method. Figure~\ref{HRdiagram_VOSA} shows the Hertzsprung--Russell diagram (HRD) built with the astrophysical parameters obtained, which are listed in Table~\ref{table:summary}. As we could not obtain a reliable fit in VOSA for star number 7028, we excluded it from our study and  focused on the remaining 250 cluster members.

\begin{table*}
\caption{Photometry uploaded to VOSA for the 251  members of M35 in our sample. This table is only partially presented. Its complete version is available electronically.}    
\label{table:photometry}      
\centering  
\small
\begin{tabular}{c c c c c c c c c c c c}     
\hline\hline        
ID & V \tablefootmark{a} & Ic \tablefootmark{b} & g \tablefootmark{c} & r \tablefootmark{c} & i \tablefootmark{c} & z \tablefootmark{c} & J \tablefootmark{d} & H \tablefootmark{d} & Ks \tablefootmark{d} \\
\hline                    
   5408 & 17.258 & 15.825 & -- & 16.91$\pm$0.05 & 16.50$\pm$0.03 & 16.13$\pm$0.03 & 14.81$\pm$0.06 & 14.17$\pm$0.06 & 13.98$\pm$0.04 \\
   5087 & 14.652 & 13.743 & 15.07$\pm$0.05 & 14.58$\pm$0.13 & 14.15$\pm$0.11 & 14.08$\pm$0.03 & 13.15$\pm$0.02 & 12.82$\pm$0.02 & 12.68$\pm$0.02 \\
   5382 & 17.136 & 15.661 & -- & 16.51$\pm$0.03 & 16.21$\pm$0.03 & 15.90$\pm$0.03 & 14.66$\pm$0.06 & 14.11$\pm$0.06 & 13.97$\pm$0.05 \\
   5194 & 15.756 & 14.607 & 16.29$\pm$0.05 & 15.41$\pm$0.05 & 15.10$\pm$0.03 & 14.88$\pm$0.03 & 13.76$\pm$0.03 & 13.27$\pm$0.03 & 13.18$\pm$0.03 \\
   7034 & -- & -- & 18.45$\pm$0.05 & 17.16$\pm$0.05 & 16.69$\pm$0.03 & 16.40$\pm$0.03 & 15.11$\pm$0.06 & 14.53$\pm$0.06 & 14.40$\pm$0.07 \\
   ... & ... & ... & ... & ... & ... & ... & ... & ... & ... \\ 
\hline                  
\end{tabular}
\tablefoot{
\tablefoottext{a}{Johnson V band magnitude taken from \cite{Ba2001b}.}
\tablefoottext{b}{Cousins I band magnitude taken from \cite{Ba2001b}.}
\tablefoottext{c}{SDSS magnitudes taken from \cite{Bo15}.}
\tablefoottext{d}{2MASS magnitudes taken from \cite{Bo15}.}
}
\end{table*}

%

   \subsection{Measuring equivalent widths and abundances}\label{EWLi_ALi}

We used the iSpec software to combine  RV-corrected versions of the spectra collected during our 1999 and 2001 campaigns. This allowed us to obtain a single spectrum for each of the stars observed during each campaign with a signal-to-noise ratio (SNR) $>$300.

We also used iSpec  to develop the pipeline to derive the EqW(Li) and Li abundance for each star. First, we used the SPECTRUM radiative transfer code described in \cite{GrayCorbally1994}, the MARCS model atmospheres \citep{Gustafsson2008}, and the \textit{Gaia}--ESO survey line list \citep{Heiter2021} to generate, for each star, a synthetic $R = 20,000$ spectrum with solar metallicity, log \textit{g} = 4.5, and the effective temperature returned by VOSA. This process was repeated several times while modifying the rotational velocity until we found the \textit{v} sin \textit{i} that best fits the science spectrum. 

Second, we determined the iron abundance for each star using  40 Fe lines between 645 nm and 680 nm together with the aforementioned stellar parameters. Once the rotational velocity and the Fe abundance have been determined, these parameters, together with the radiative transfer code, model atmospheres, and line list previously mentioned, were used to derive the EW of the Li doublet at 670.78 nm as well as the local thermodynamic equilibrium (LTE) Li abundance. Both quantities were measured three times for each star while modifying slightly the continuum value, and their uncertainties were calculated as the standard deviation of those measurements.

   \begin{figure*}
   \centering
   \includegraphics[width=9cm]
   {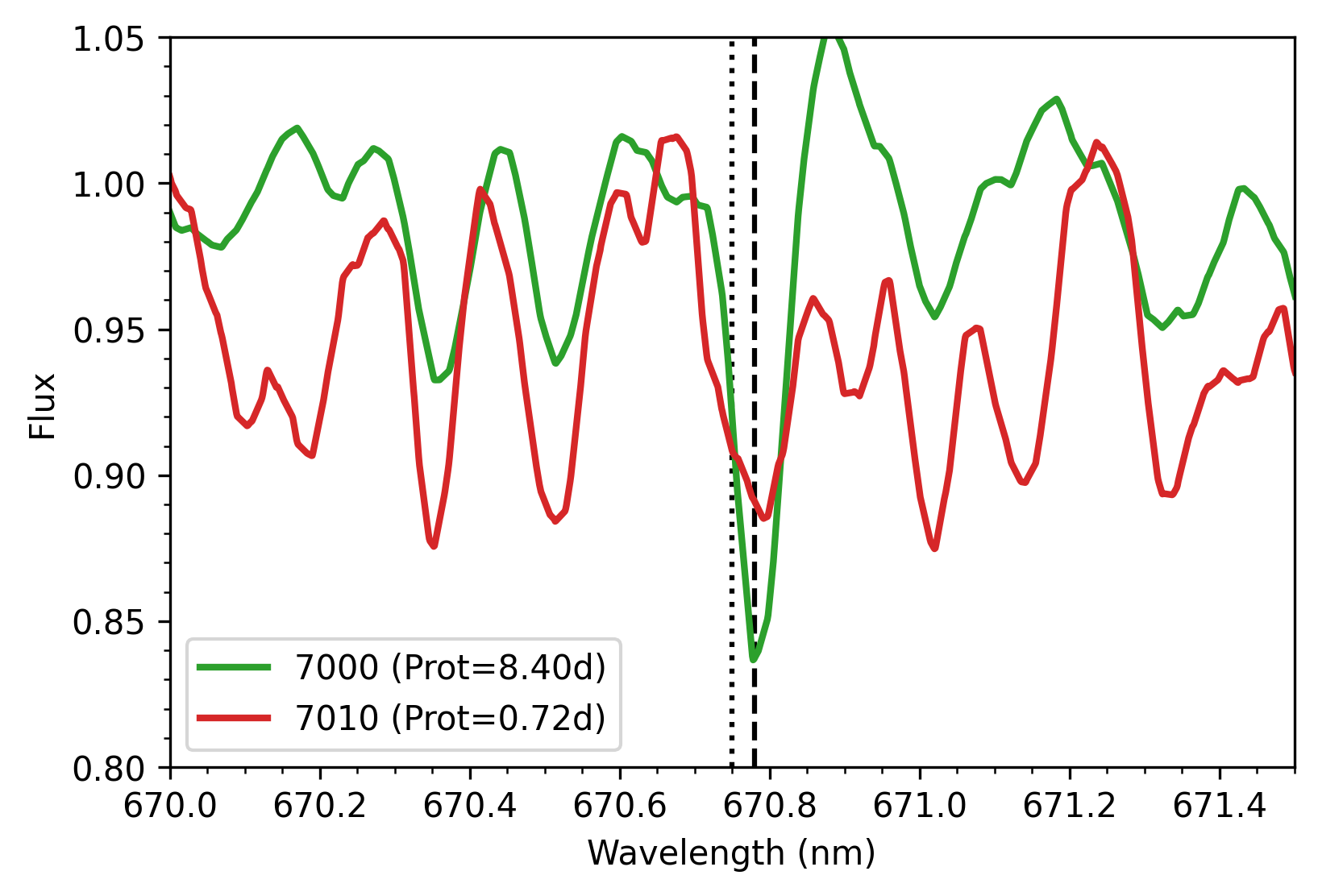}
   \includegraphics[width=9cm]
   {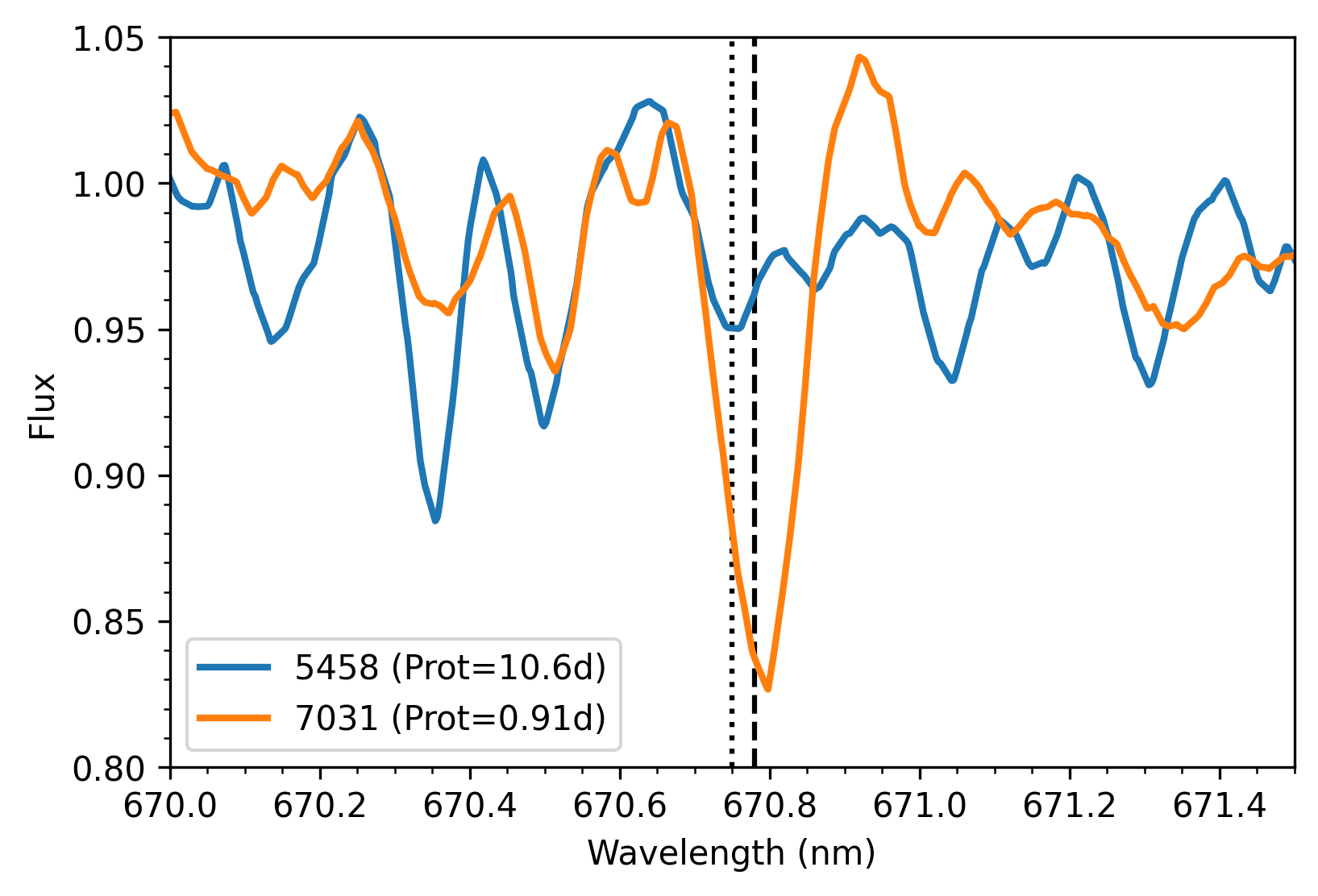}
   \caption{Blended feature around 670.78 nm for four K stars. \textit{Left---} Small region of the spectrum for two mid K stars with different rotation periods. \textit{Right---} Small region of the spectrum for two late K stars with different rotation periods. In both panels, the dashed black line shows the position of the Li doublet at 670.78 nm, while the dotted black line indicates the position of the Fe I line at 670.75 nm. Both features are blended independently of the spectral type and the rotation period of the star.}
              \label{LiI_line_blended}%
    \end{figure*}

However, as the resolution of the spectra is not high enough to separate the Li doublet at 670.78 nm and the Fe I line at 670.75~nm, both features are blended (see  Figure~\ref{LiI_line_blended}, where a small region around the Li doublet is shown for four K stars). To remove the contribution of this iron line, we have tried two different approaches. 

First, we built a synthetic spectrum employing a modified version of the \textit{Gaia}--ESO line list where all the Li lines had been removed. The SNR estimated for the science spectrum was used to add errors to the synthetic one for each star. The EW corresponding to the Fe I line at 670.75~nm was then subtracted from the EW of the blend, resulting in a measurement of the EW of the Li doublet. The uncertainty in EW(Li) width was derived from the quadratic sum of the uncertainty in the blend and the uncertainty in the EW of the Fe I line. 

Second, we followed the approach of \cite{Arancibia2020}: we derived (B-V) colours for our stars from the dereddened \textit{Gaia} (BP - RP) colours, and we applied the empirical relation derived in \cite{soderblom1993}. To obtain intrinsic (B-V) colours for the targets observed in these campaigns we linearly interpolated through the  \cite{PecautMamajek2013} (B-V)$_0$ -- (BP-RP)$_0$ relationship. Finally, we calculated the contribution of the the Fe I line at 670.75 nm as \textit{EW(Fe)} = 20 \textit{(B-V)$_0$} $-$ 3 m$\AA$. 

\begin{figure*}
   \centering
   \includegraphics[width=9cm]
   {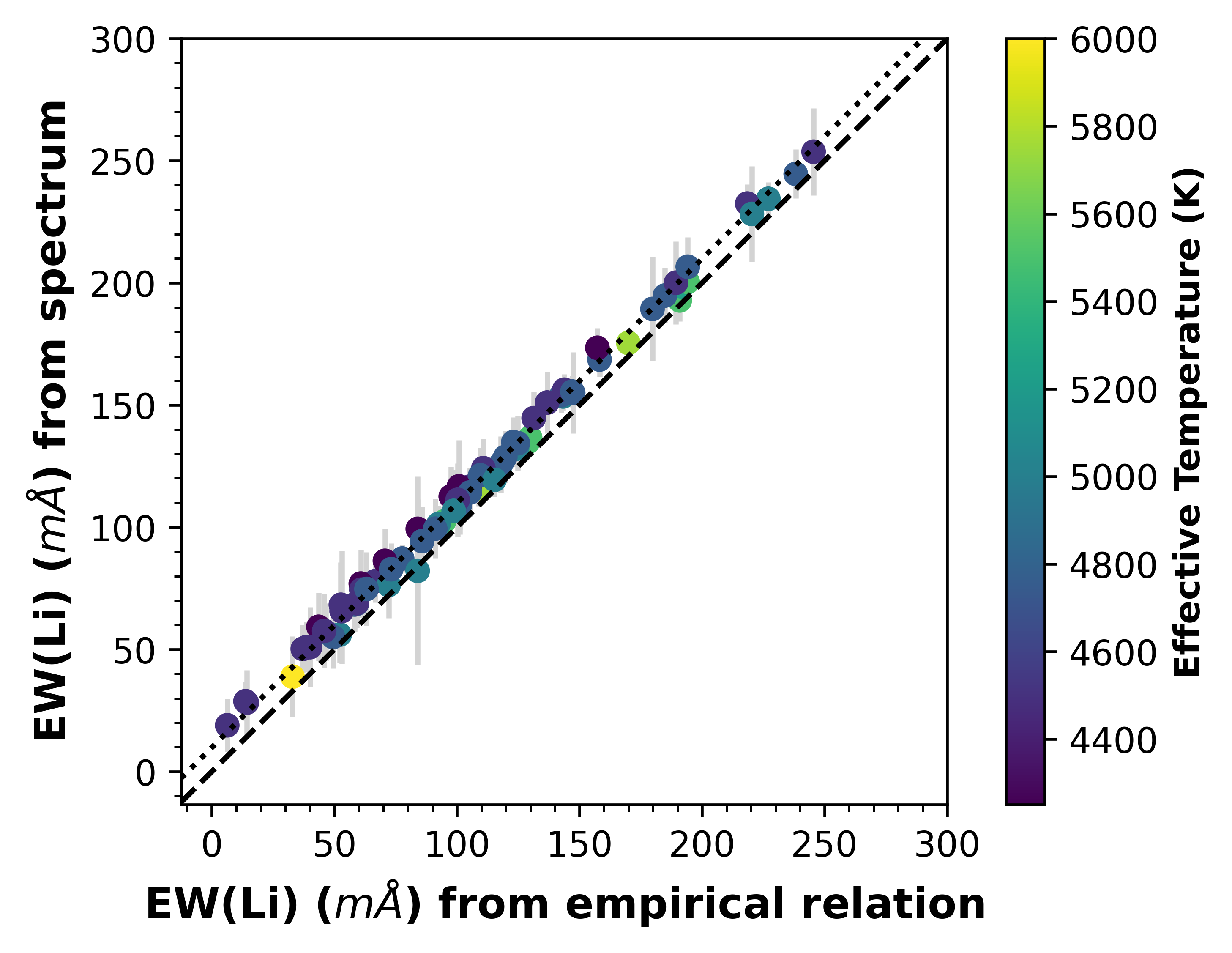}
   \includegraphics[width=9cm]
   {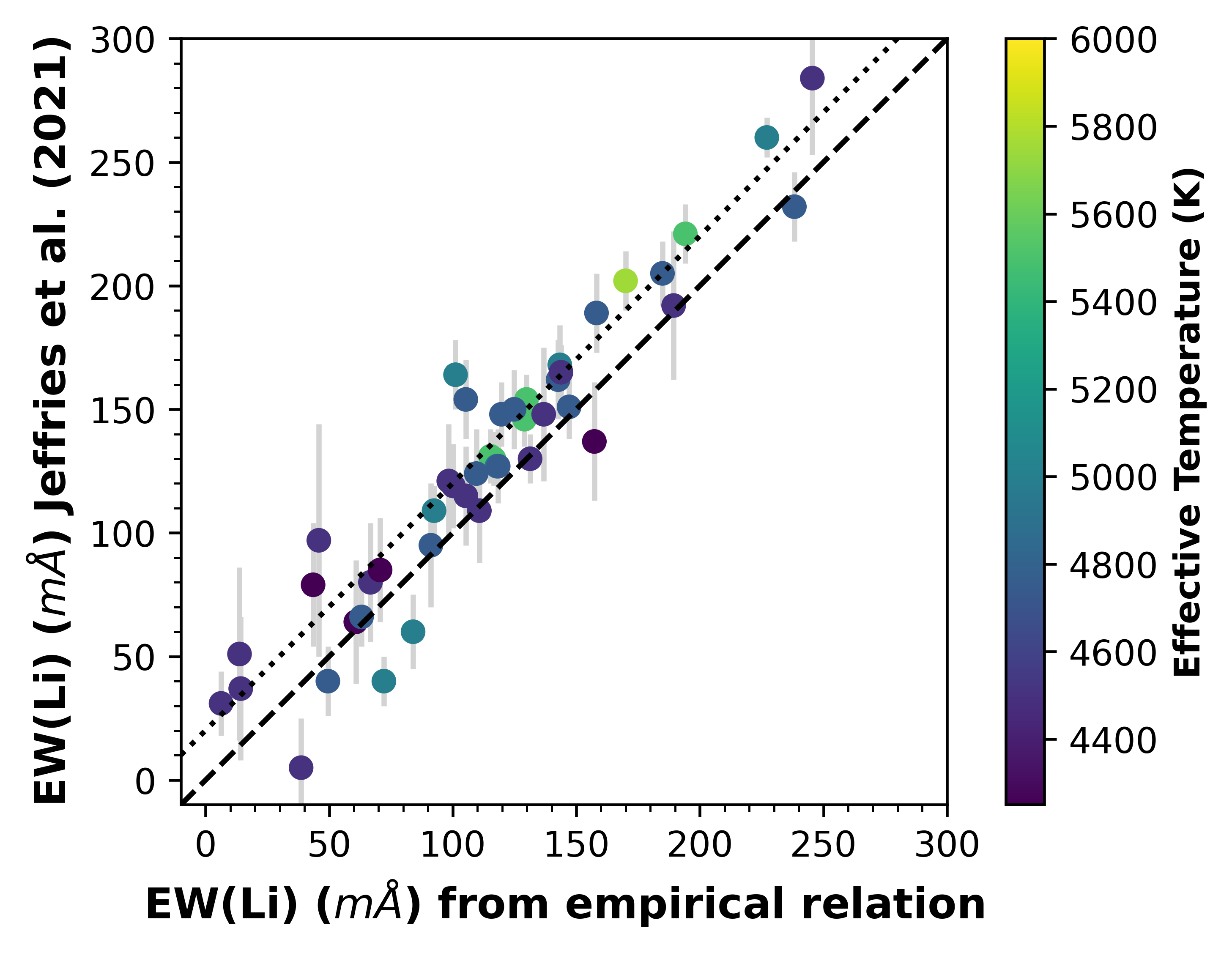}
   \caption{Comparisons between the Li EWs derived by applying the empirical relation published in \cite{soderblom1993} to the M35 spectra collected in 1999 December and 2001 February  (x axes) and the EWs obtained by other methods for the same objects (y axes). \textit{Left---}Comparison to Li EWs obtained using synthetic spectra without Li generated with the SPECTRUM code and MARCS model atmospheres. \textit{Right---}Comparison to Li EWs published in \citet{Je21}. In both panels, the dashed black line shows the one-to-one correspondence, while the dotted  line indicates the median of the differences. The colour of each symbol indicates the effective temperature derived for the star. In both cases, the EWs indicated on the y axes are, on average, larger than the ones shown on the x axes.}
              \label{EWLi_TwoMethods}%
    \end{figure*}

The left panel on Figure \ref{EWLi_TwoMethods} shows the comparison between the Li EWs obtained by these two approaches. The median of the differences is around 10 m$\AA$, as indicated by the dotted black line, and the spread around this value is small. 

The right panel of Figure \ref{EWLi_TwoMethods} shows the comparison between the Li EWs published in \citet{Je21} and the values obtained by employing the \cite{soderblom1993} relation. \citet{Je21} measured the Li EWs by comparing the science spectra with synthetic spectra without Li generated with the MOOG software \citep{sneden2012} and solar-metallicity Kurucz model atmospheres \citep{kurucz1992}. The median of the differences is around 20 m$\AA$ in this case, as indicated by the dotted black line, and the spread around this line is broader than the one shown in the left panel. Since the equivalent widths shown in the left panel have been derived from the same spectra, unlike the ones shown in the right panel, a broader spread in the latter is not surprising.

\begin{figure*}
   \sidecaption
   \includegraphics[width=12cm]{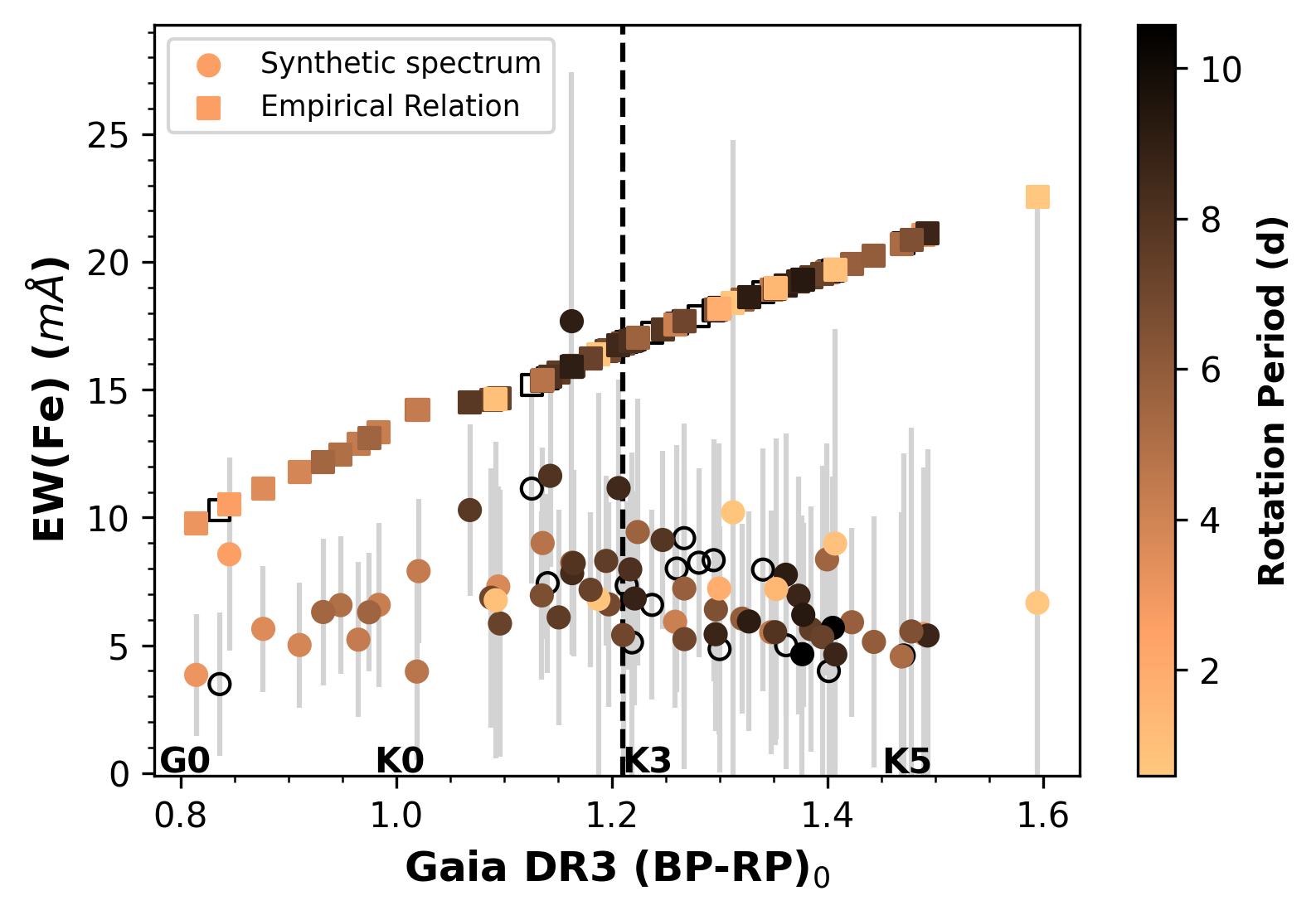}
   \caption{EWs measured for the Fe I line at 670.75 nm in the synthetic spectra and those derived from the empirical law proposed by \cite{soderblom1993}. \textit{Gaia} colours have been dereddened using \textit{E(BP - RP)} = 0.415 \textit{$A_{V}$}, following \citet{curtis2020}. The colour of each symbol indicates the rotation period of the star. The \cite{soderblom1993}  relation was derived from spectra of stars on the left side of the vertical dashed line. The EWs obtained for stars on the right side of the line are therefore obtained by extrapolating that relation to a cooler range of temperatures.}
              \label{EWFe_colour}
    \end{figure*}

In Figure~\ref{EWFe_colour}, we plot the iron equivalent widths obtained from the \cite{soderblom1993} empirical relation and those measured in the synthetic spectra, confirming a $\approx$5 m$\AA$ difference between G0 and early K spectral types, which gets larger than 10 m$\AA$ for stars cooler than (BP-RP)$_0$ = 1.2. Since \cite{soderblom1993} derived their relation from the spectra of inactive, old field stars with spectral types earlier than K3, a large difference for cooler dwarfs is not surprising. The smaller differences observed in hotter stars could have a methodological or an intrinsic origin. It is possible that the empirical relation proposed by \cite{soderblom1993} takes into account fainter contributions from other elements different from iron that are not included in the synthetic spectra. However, given the completeness of the line list employed to generate the synthetic spectra, this explanation is unlikely. An alternative explanation could be that the \cite{soderblom1993} relation is useful for solar metallicity clusters, like the Pleiades, but overestimates the contribution of the Fe I line at 670.75 nm to the blended feature for subsolar metallicity clusters, such as M35. Consequently, we used the Li EWs obtained using synthetic spectra to remove the contribution of the Fe I line for our analysis.

Finally, we estimated the departures from LTE for the Li abundances derived from the non-LTE calculations published in \cite{Lind2009}. To do so, we split our stars in bins of effective temperature and linearly interpolated through \cite{Lind2009} non-LTE correction--A(Li) LTE relationship.\footnote{We selected the rows in table 2 of \cite{Lind2009} for a solar metallicity and microturbulence $\xi$$_t$ = 1.0 \textit{km s$^{-1}$}.}

Table~\ref{table:ALi} gives the derived Li EWs, the Fe EWs obtained with the two methods described above, as well as the LTE and non-LTE Li abundances for our stars.

\begin{table*}
\caption{Blended Li and Fe EWs, Fe EWs, Li EWs, and Li LTE and non-LTE abundances for the 110 members of M35 observed in 1999 December and 2001 February. This table is only partially presented. Its complete version is available electronically.}             
\label{table:ALi}      
\centering          
\begin{tabular}{c c c c c c c}     
\hline\hline        
ID & \textit{EW(Li+Fe)} & \textit{EW(Fe)} \tablefootmark{a} & \textit{EW(Fe)} \tablefootmark{b} & \textit{EW(Li)} \tablefootmark{a} & \textit{A(Li)$_{LTE}$} & \textit{A(Li)$_{NLTE}$} \\
 & (m$\AA$) & (m$\AA$) & (m$\AA$) & (m$\AA$) &  & \\
\hline                    
   5088 & 119.2$\pm$6.5 & 3.5$\pm$2.8 & 10.3 & 115.7$\pm$7.1 & 2.88$\pm$0.04 & 2.85 \\
   5103 & 201.4$\pm$7.6 & 8.6$\pm$3.8 & 10.5 & 192.9$\pm$8.5 & 2.98$\pm$0.03 & 2.92 \\
   5139 & 127.1$\pm$5.2 & 5.0$\pm$2.5 & 11.8 & 122.0$\pm$5.7 & 2.66$\pm$0.03 & 2.68 \\
   5148 & 181.1$\pm$4.7 & 5.6$\pm$2.5 & 11.1 & 175.5$\pm$5.3 & 3.22$\pm$0.03 & 3.06 \\
   5183 & 107.6$\pm$5.1 & 5.2$\pm$3.0 & 12.9 & 102.3$\pm$5.9 & 2.57$\pm$0.03 & 2.59 \\
   5295 & $\leq$14.3 \tablefootmark{c} & -- & -- & -- & -- & -- \\
   ... & ... & ... & ... & ... & ... & ... \\ 
\hline                  
\end{tabular}
\tablefoot{
\tablefoottext{a}{Value obtained from the synthetic spectrum without Li.}
\tablefoottext{b}{Value obtained from the empirical relation of \cite{soderblom1993}.}
\tablefoottext{c}{Upper limit. Li abundances and Fe EWs are not provided for these cases.}
}
\end{table*}

	\section{Results} \label{sec:results}

   \subsection{The metallicity of M35}\label{FeH_M35}

Previous WIYN/Hydra spectroscopic studies of Li in M35 quoted different values for its metallicity. \citet{Ba01a} derived [Fe/H] = $-$0.21$\pm$0.10 by measuring Fe EWs for a set of single G members of the cluster,  whereas \cite{steinhauer2004} set [Fe/H]~=~$-$0.143$\pm$0.014 when studying Li depletion in the cluster's F dwarfs. On the other hand, \citet{At18} measured the EW of the Ca I line at 671.77 nm and obtained results compatible with solar metallicity.  \citet{Je21} assumed a solar metallicity for their analysis, but discussed the impact a lower metallicity could have on their results.

As discussed in \S\ref{EWLi_ALi}, we used 40 iron lines to derive the abundance of this element for each of the M35 members observed in 1999 December and 2001 February. This list includes 38 Fe I transitions and two Fe II transitions. The results obtained for those single stars with the highest probability of belonging to M35 were employed to derive our own value for the metallicity for this cluster.

\begin{figure}
   \resizebox{\hsize}{!}{\includegraphics{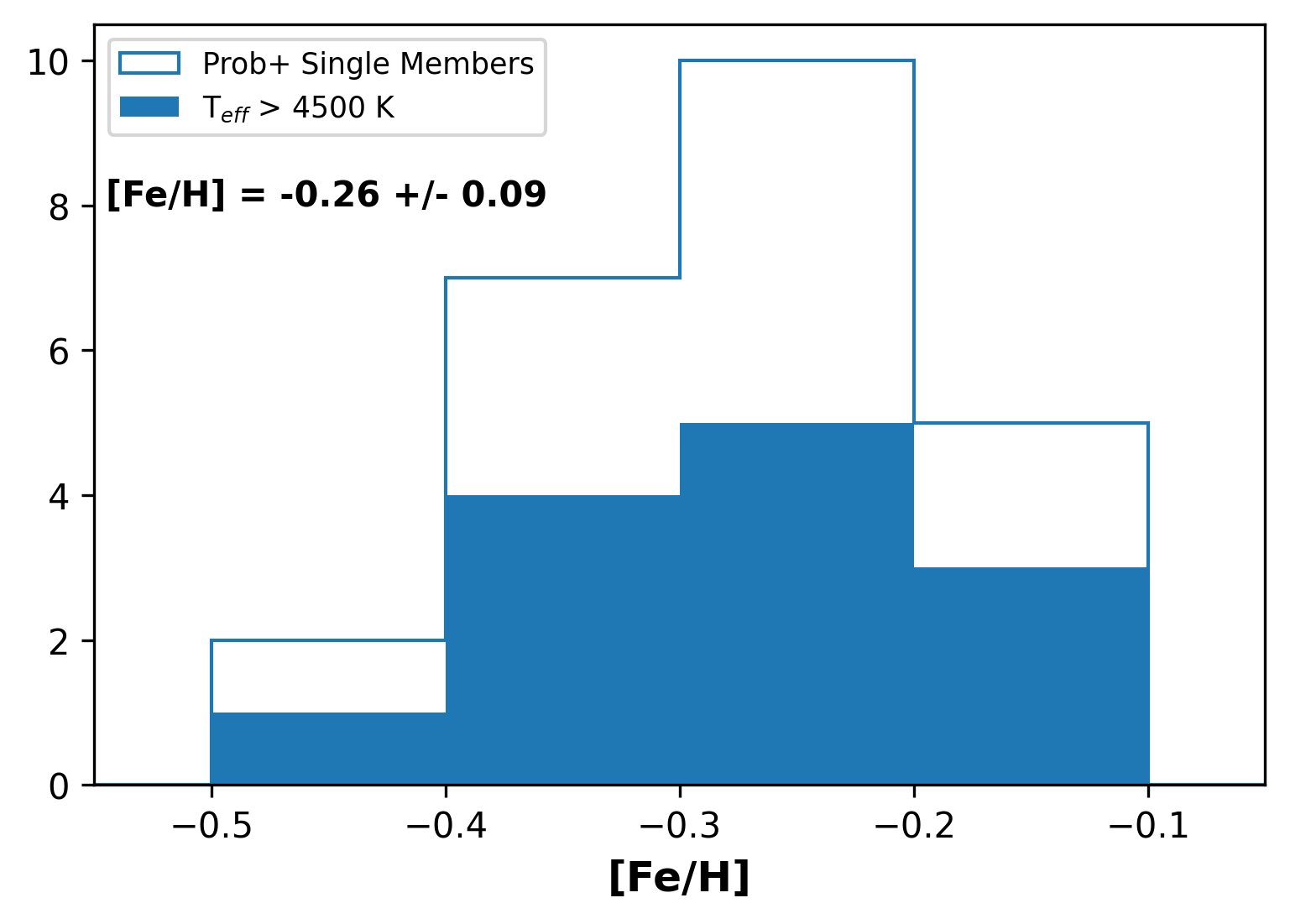}}
   \caption{Metallicities obtained for the single stars observed in 1999 December and 2001 February that have the highest probability of belonging to M35 (\textit{Prob+} members). The unfilled histogram represents the distribution of metallicities obtained for all of them, whereas the filled histogram only includes the hottest stars in this sample. We used the values corresponding to the hottest ones to derive a metallicity of [Fe/H] = $-$0.26$\pm$0.09 for M35.}
              \label{FeH_hist}%
    \end{figure}

As shown in Figure \ref{FeH_hist}, our results clearly point to a subsolar metallicity for M35. We have only considered the hottest stars (\textit{T$_{eff}$} > 4500 K) in the set previously described, obtaining an average metallicity of [Fe/H]~=~$-$0.26$\pm$0.09, where the standard deviation of the values has been used to indicate the uncertainty. This result is fully consistent with the values published in \citet{Ba01a} and \cite{steinhauer2004}.

%

   \subsection{The Li-rotation connection in M35}

\begin{table*}
\caption{Rotation period, Li EW, effective temperature, luminosity, membership class and binary status for 250 M35 members. This table is only partially presented. Its complete version is available electronically.}
\label{table:summary}
\centering          
\begin{tabular}{c c c c c c c c c}     
\hline\hline        
ID & Membership & Binary \tablefootmark{a} & \textit{T$_{eff}$} & \textit{L} & \textit{EW(Li)} & \textit{EW(Li)} \tablefootmark{b} & Prot \tablefootmark{c} & Prot \tablefootmark{d} \\
 & Class &  & (K) & (L$_{\odot}$) & (m$\AA$) & Source & (d) & Source \\
\hline                    
   5408 & Poss+ & 0 & 4500$\pm$125 & 0.21$\pm$0.04 & 291 & B01 & 0.74 & M09 \\
   5087 & Poss+ & 0 & 6000$\pm$125 & 1.57$\pm$0.79 & 101.7 & AT18 & 2.20 & SF20 \\
   5382 & Prob & 0 & 4750$\pm$125 & 0.26$\pm$0.05 & 187 & Je21 & 3.08 & M09 \\
   5194 & Poss & 1 & 5000$\pm$125 & 0.68$\pm$0.12 & 153.6 & This Work & 4.70 & M09 \\
   7034 & Prob & 0 & 4500$\pm$125 & 0.16$\pm$0.03 & $\leq$ 29.9 & This Work & 9.71 & M09 \\
   5107 & Poss+ & 0 & 6000$\pm$125 & 1.38$\pm$0.26 & 94 & B01 & 2.51 & This Work \\
   ... & ... & ... & ... & ... & ... & ... & ... & ... \\ 
\hline                  
\end{tabular}
\tablefoot{
\tablefoottext{a}{1 = likely binary, 0 = single star.}
\tablefoottext{b}{B01 is \cite{Ba01a}, AT18 is \cite{At18}, and Je21 is \cite{Je21}.}
\tablefoottext{c}{-99 = No rotation period available for this source.}
\tablefoottext{d}{M09 is \citet{Meibom2009}, SF20 is \citet{SF20}, N15 is \citet{N15}, and L16 is \citet{Libralato2016}.}
}
\end{table*}
 
\begin{figure*}
	\sidecaption
	\includegraphics[width=12cm]{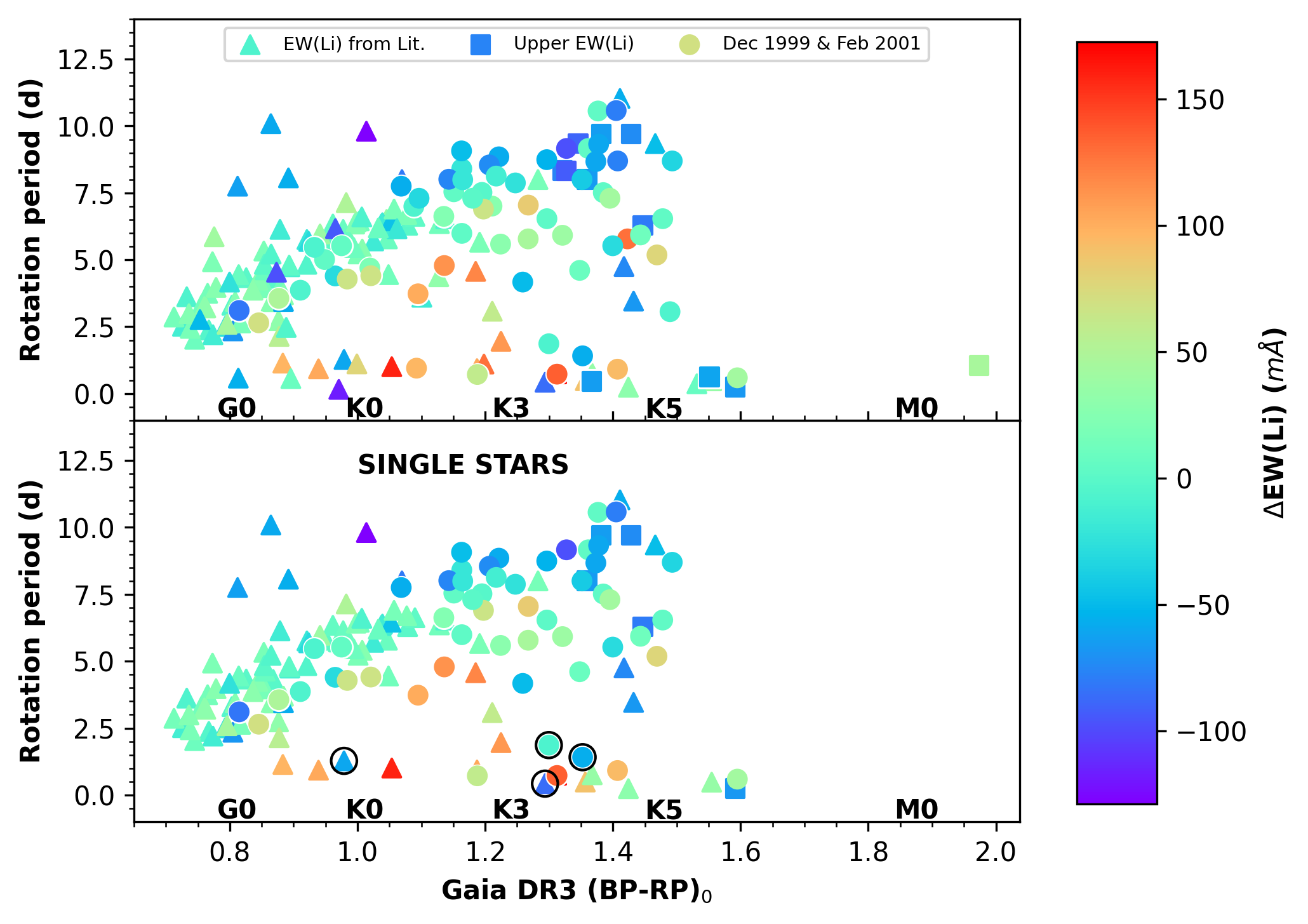}
	\caption{Distribution of rotation periods for the M35 members in our sample. The colour of each symbol indicates the deviation of the EW(Li) obtained for that star from a third order polynomial fit to EW(Li) vs.~\textit{(BP - RP)$_0$}. Circles indicate  the M35 members observed in 1999 December and 2001 February whose Li EWs have been derived as discussed in section \ref{EWLi_ALi}, while triangles are those M35 members whose Li EWs have been taken from \citet{Ba01a}, \citet{At18}, or \citet{Je21}. Squares denote those stars observed in 1999 December and 2001 February for which we could only derive upper limits for the Li EWs. \textit{Top---}Our complete set of M35 members with rotation periods. \textit{Bottom---}Only those sources classified as single stars are shown. The four outliers that do not follow the general trend are circled in black.}
		\label{Prot_Teff_M35}
	\end{figure*}

Figure \ref{Prot_Teff_M35} is a color--period diagram in which we indicate the relative Li abundances of the stars in M35; the top panel includes all of the cluster members, whereas in the bottom panel we have removed candidate and known binary systems. Table~\ref{table:summary} includes the stellar parameters, rotation periods, and Li EWs for these stars. The most striking result is that M35 fast rotators of G and K spectral types are Li-rich compared with slow rotators of similar effective temperature. This result is consistent with the pattern described in \citet{Ba01a}, \citet{At18} and \citet{Je21} for the M35 open cluster as well as with the trend found for other stellar associations \citep[see][and references therein]{Bouvier2020}.

Few studies have addressed the influence of a stellar companion on the Li abundance. \cite{Martin2002} studied the abundance of this element in wide binaries with solar-type twin components, finding large differences between the components for a significant part of the sample. \cite{Barrado1996} demonstrated that tidally locked binaries in the Hyades open cluster are richer in Li in comparison with single stars. The purpose of the present article is not to characterize Li in binary systems, but it is worth highlighting that the difference in Li EWs between fast and slow G and K rotators is more evident in the bottom panel of Figure \ref{Prot_Teff_M35}, where multiple systems are not included. 

The M0 fast rotator shown in the top panel deserves special attention. This star is a spectroscopic binary with identification number 5517 which was observed in 1999 December. We have measured EW(Li) $\leq$ 24.1 m$\AA$ for this object, which is a value very similar to the EWs derived by \cite{BouvierPleiades2018} and \cite{BarradoPleiades2016} for Pleiades members of similar spectral type and rotation period.

Four outliers, which are circled in the bottom panel of Figure~\ref{Prot_Teff_M35}, do not follow the trend described above. The hottest, whose Li EW has been taken from \citet{At18}, has the identification number 5181 \citep[46015 in][]{At18}. Its rotation period has been taken from \citet{Meibom2009}, but very similar periods are considered ambiguous blends in \citet{SF20} and \citet{Libralato2016}, so the rotation period of this source may be different from the one we have assigned it in Table \ref{table:Prot}. This source is also considered as an outlier in \citet{At18} (see section 4 of that work). 

Two fast rotators, 7049 and 7053, observed in 1999 December and 2001 February have Li EWs that are surprisingly low. The membership of star 7049 is doubtful since both the probability provided by Clusterix 2.0 and the radial velocity measured for this star indicate it does not belong to M35. Consequently, the membership probability assigned in \cite{Bo15} to this source may be wrong. By contrast, star 7053 has the highest probability of belonging to M35. However, only \cite{Meibom2009} provides rotation period for this star, and comparing their result with those of others is therefore not possible. Both sources were studied by \cite{Je21}, who obtained Li EWs compatible with the ones shown in Table \ref{table:ALi}. 

Finally, the equivalent width of the remaining outlier, star 5290, has been taken from \citet{Je21}. In contrast to the other sources, there is no reason to think that the rotation period assigned to this star is mistaken, nor that it is not member of M35. Follow-up observations of this star are needed to understand the reason why it is Li-poor in comparison with the other fast rotators. In any case, it is clear that the aforementioned connection between Li and rotation holds.

   \subsection{Comparison with other clusters}\label{M35_Pleiades_M34}

\begin{figure*}
	\sidecaption
	\includegraphics[width=12cm]{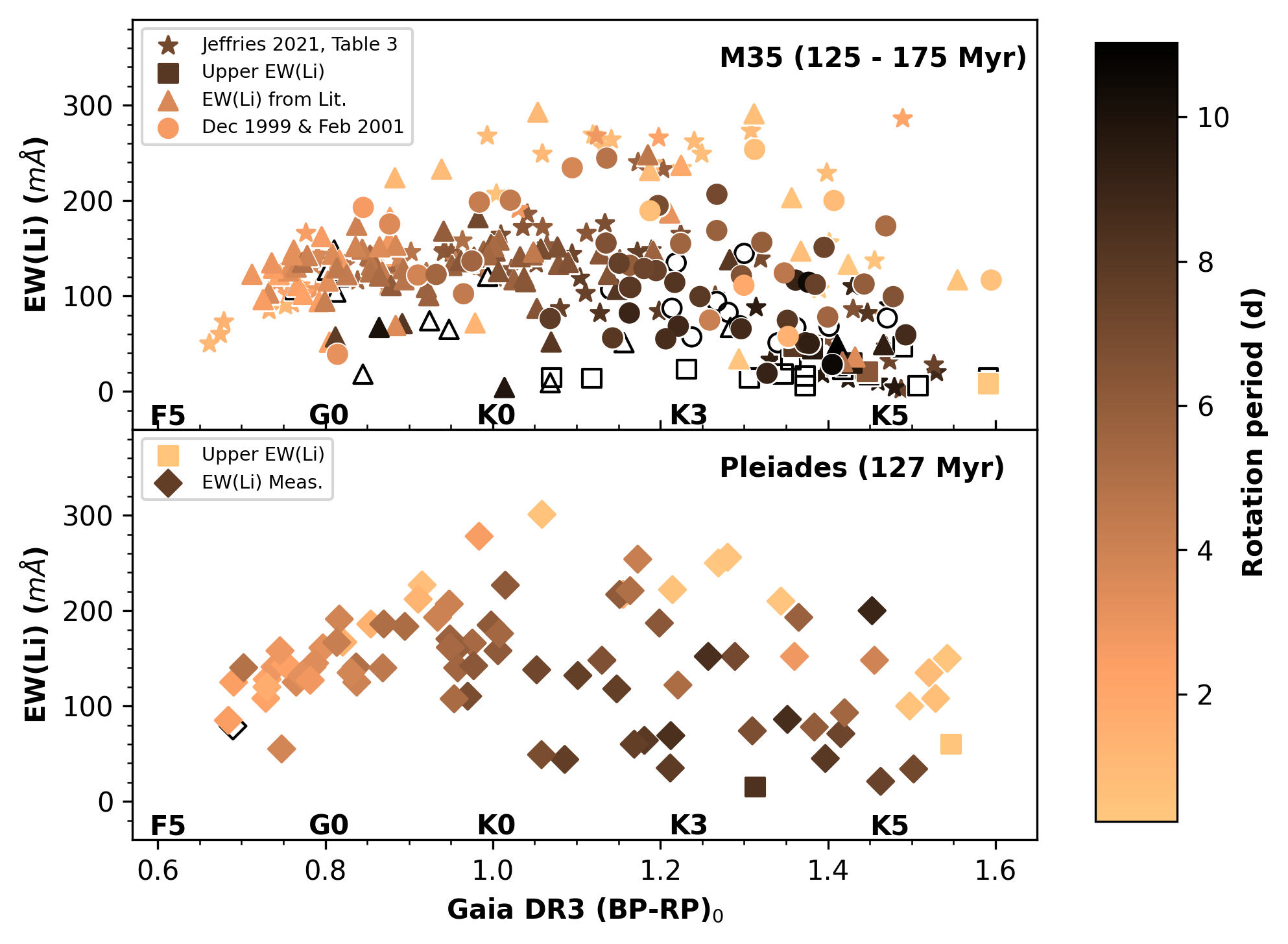}
	\caption{Li EWs vs.~\textit{Gaia} colours for single members of M35 and the Pleiades. The symbols are colour-coded by the rotation period of the star. The empty symbols are M35 members without a measure rotation period. \textit{Top---}M35 single stars. Circles, triangles, and squares have the same meaning as in Figure \ref{Prot_Teff_M35}. We use star symbols to represent the  single stars  analysed by \citet{Je21} that do not have counterparts in our sample. \textit{Bottom---}Pleiades single stars taken from \cite{BarradoPleiades2016} and  \cite{BouvierPleiades2018}. Note that we have chosen the Li EWs and rotation periods published in \cite{BouvierPleiades2018} for all of the stars that are included in both studies. We have only plotted those Pleiades members whose \textit{Gaia} colours are within the range covered by the M35 sample shown in the top panel. \textit{Gaia} colours have been dereddened following \citet{curtis2020} with \textit{A$_V$} = 0.12.}
		\label{EWLi_Teff_M35_Pleiades}
	\end{figure*}

\begin{figure*}
	\sidecaption
	\includegraphics[width=12cm]{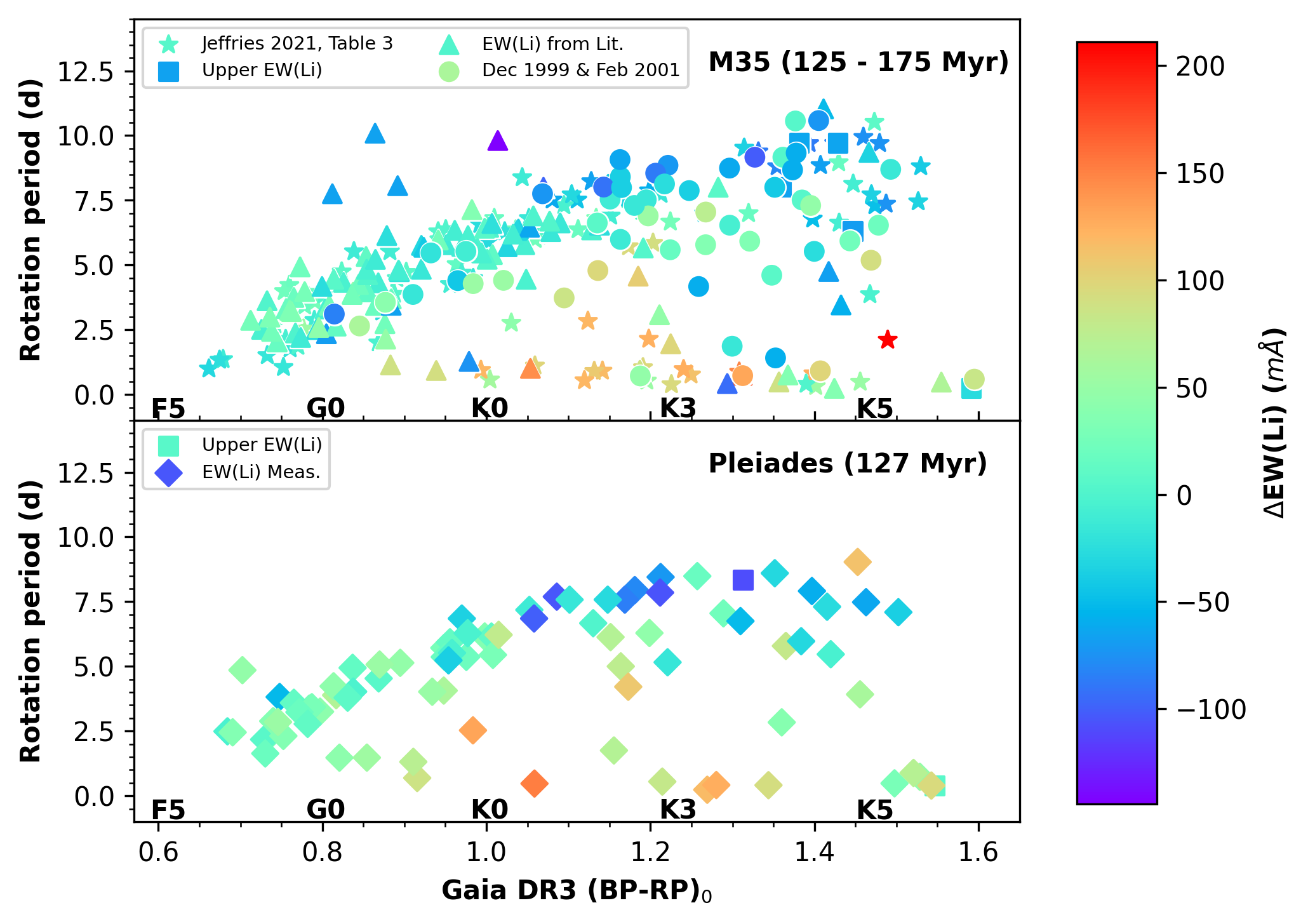}
	\caption{Rotation periods vs.~\textit{Gaia} colours for single members of M35 and the Pleiades. The colour of each symbol indicates the deviation of EW(Li) from a third-order polynomial fit to EW(Li) vs.~\textit{(BP - RP)$_0$} for the M35 sample shown in the upper panel. \textit{Top---} M35 single stars. Circles, triangles, and squares have the same meaning as in Figure \ref{Prot_Teff_M35} whereas stars account for the M35 single sources analysed in \citet{Je21} which do not have counterpart in our sample. \textit{Bottom---} Pleiades single stars taken from \cite{BarradoPleiades2016} and  \cite{BouvierPleiades2018}. Note that we have chosen the Li EWs and rotation periods published in \cite{BouvierPleiades2018} for all the sources which have counterpart in both studies. We have only plotted those Pleiades members whose \textit{Gaia} colours are within the range covered by the M35 sample shown on the top panel. \textit{Gaia} colours have been dereddened following \citet{curtis2020} with \textit{A$_V$} = 0.12.}
		\label{Prot_Teff_M35_Pleiades}
	\end{figure*}

We have included in Figure \ref{EWLi_Teff_M35_Pleiades} and Figure \ref{Prot_Teff_M35_Pleiades} the single stars studied by \citet{Je21} that do not have counterparts in our sample (top panels), as well as equivalent data for the Pleiades for comparison (bottom panels). \cite{Galindo2022} recently provided a robust value of 127.4$^{+6.3}_{-10}$ Myr for the age of the Pleiades by employing the Li depletion boundary technique described in \cite{Soderblom2014}, \citet[][\citeyear{rebolo1996}]{rebolo1992}, and \citet[][\citeyear{stauffer1999}]{stauffer1998}. Although the age of M35 is affected by a much higher uncertainty, given the values provided by \citet{Je21} and \citet{At18} (140$\pm$15 Myr and 150$\pm$25 Myr, respectively) both clusters could be coeval. In fact, \cite{Abdelaziz2022} obtained an age of 126 Myr for M35 by using the Padova isochrones of \cite{PadovaIso} to fit the turnoff point in several color--magnitude diagrams. We assumed that the age of M35 is between 125 Myr and 175 Myr, the latter being the value provided by \citet{Ba01a}. 

The metallicity of the Pleiades is very close to solar. \cite{Boesgaard1990} derived [Fe/H] = $-$0.034$\pm$0.024 from a sample of F dwarfs whereas \cite{Gebran2008} obtained [Fe/H] = 0.06$\pm$0.02. Given the metallicity derived for M35 by \citet{Ba01a}, \cite{steinhauer2004}, and this work, M35 is metal poor in comparison with the Pleiades. Despite this difference in metal content and a possible difference in age, the trends shown in Figures~\ref{EWLi_Teff_M35_Pleiades} and \ref{Prot_Teff_M35_Pleiades} for M35 and the Pleiades are very similar. In both cases, fast rotators are richer in Li than slow rotators of the same spectral type. This connection is more evident for K stars than for G stars as the spread is larger in that range of effective temperatures.

%

\begin{figure*}
	\sidecaption
	\includegraphics[width=12cm]{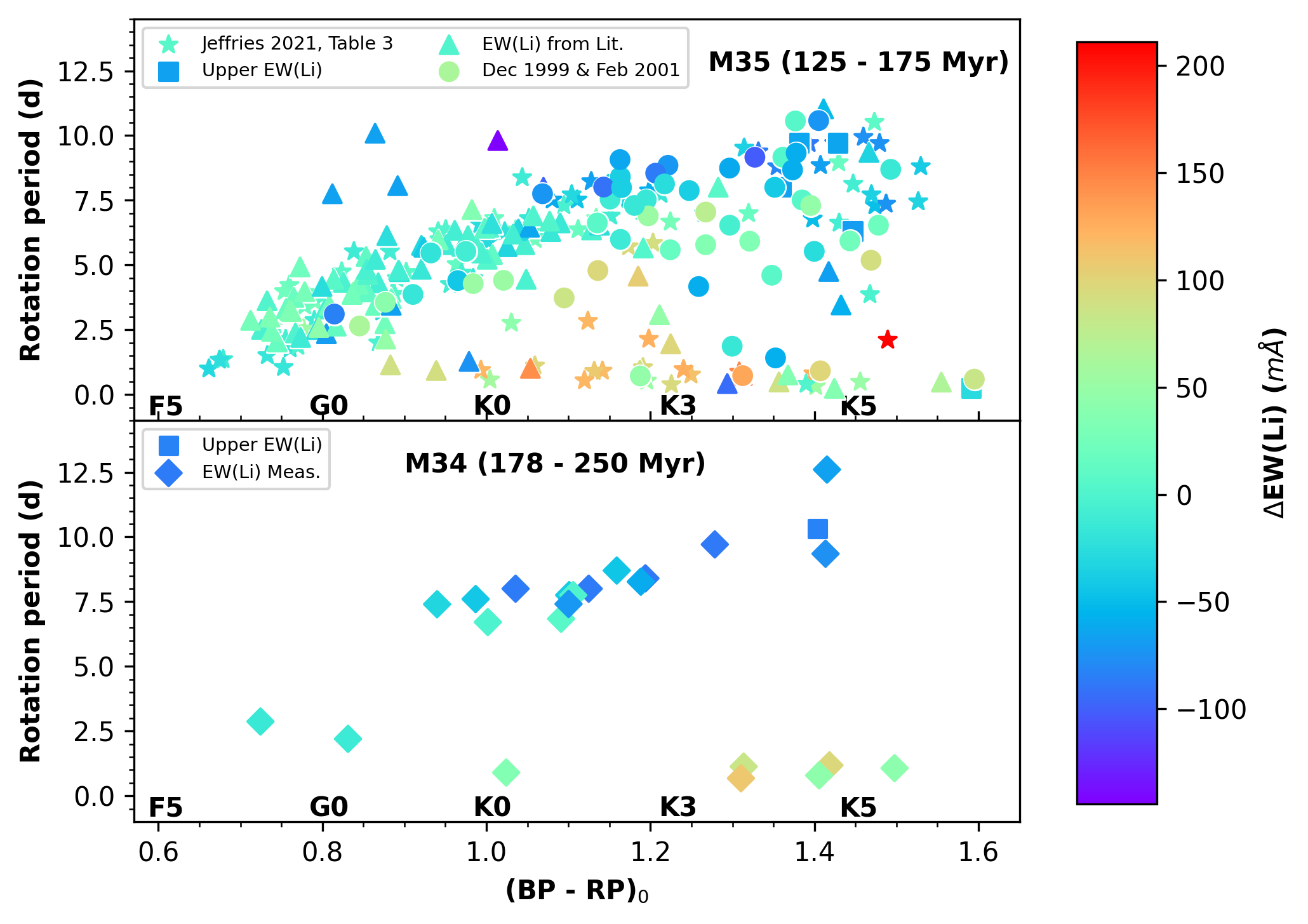}
	\caption{Rotation periods vs.~\textit{Gaia} colours for single members of M35 and M34. The colour of each symbol indicates the deviation of EW(Li) from a third order polynomial fitted to EW(Li) vs \textit{(BP - RP)$_0$} for the M35 sample shown in the upper plot. \textit{Top---}M35 single stars. Circles, triangles, squares, and stars have the same meaning as in Figure \ref{Prot_Teff_M35_Pleiades}. \textit{Bottom---}M34 single stars from \cite{Gondoin2014}. \textit{Gaia} colours have been dereddened following \citet{curtis2020} with \textit{A$_V$} = 0.217.}
		\label{Prot_Teff_M35_M34}
	\end{figure*}

We also compared our results for M35 to those for M34. \cite{Ianna1993} estimated an age of 250 Myr for M34 through isochrone fitting, whereas \cite{Meynet1993} derived an age of 178 Myr for this cluster employing the same technique. \cite{Schuler2003} derived [Fe/H] = 0.07$\pm$0.04 from high resolution spectra of five G-K dwarf members of the cluster. Since M34 is older than M35 and has solar metallicity, as do the Pleiades, we have compared our sample of M35 single dwarfs of G-K spectral type with the one presented in \citet{Gondoin2014} for M34 (Figure \ref{Prot_Teff_M35_M34}). Although the similarities between both distributions are more evident in Figure \ref{Prot_Teff_M35_Pleiades}, as the Pleiades sample is more similar in size to that for M35, a similar trend can be observed for M34 K dwarfs: the fast rotators are richer in Li than the slow ones. 

\begin{figure*}
   \centering
   \includegraphics[width=9cm]
   {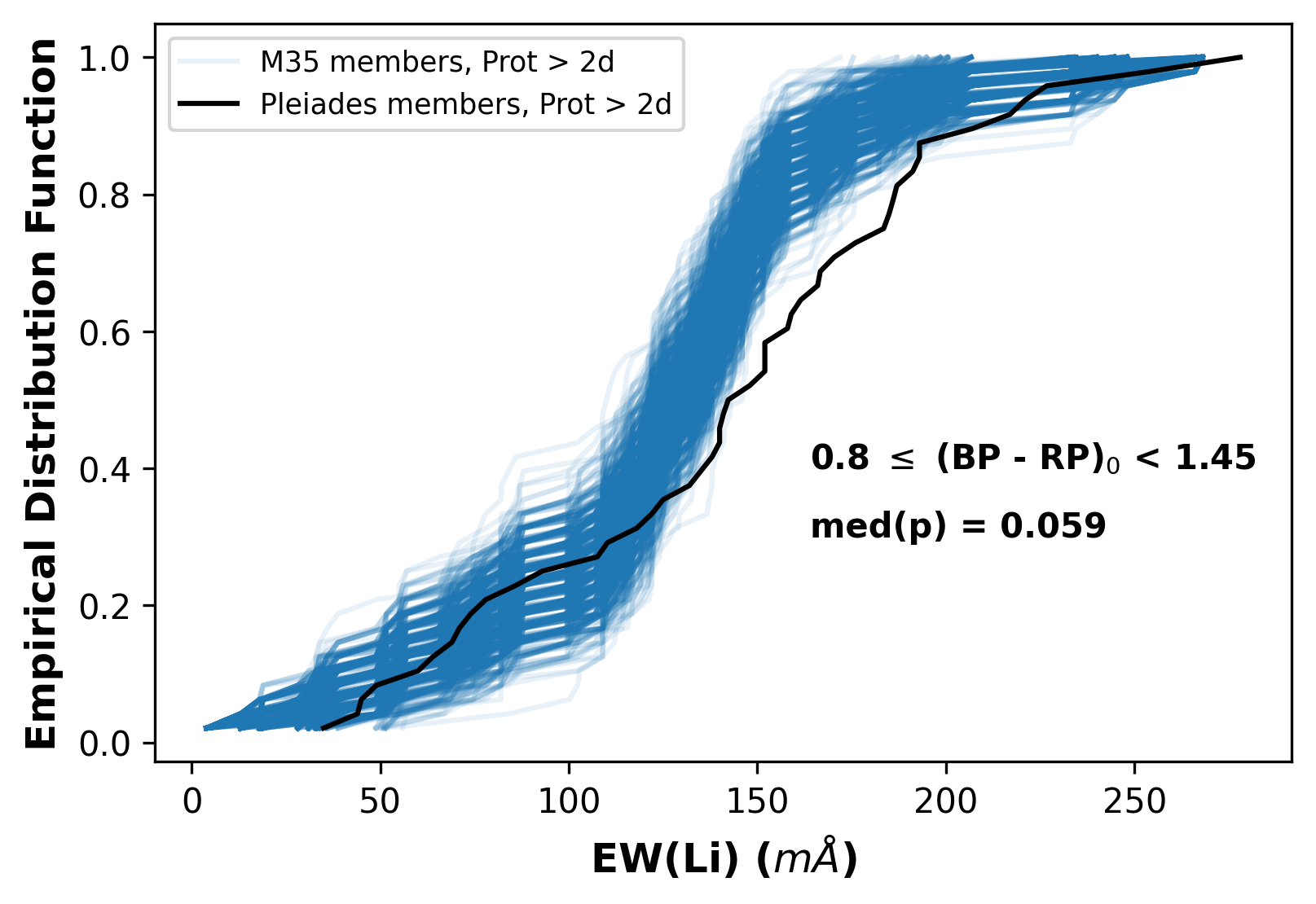}
   \includegraphics[width=9cm]
   {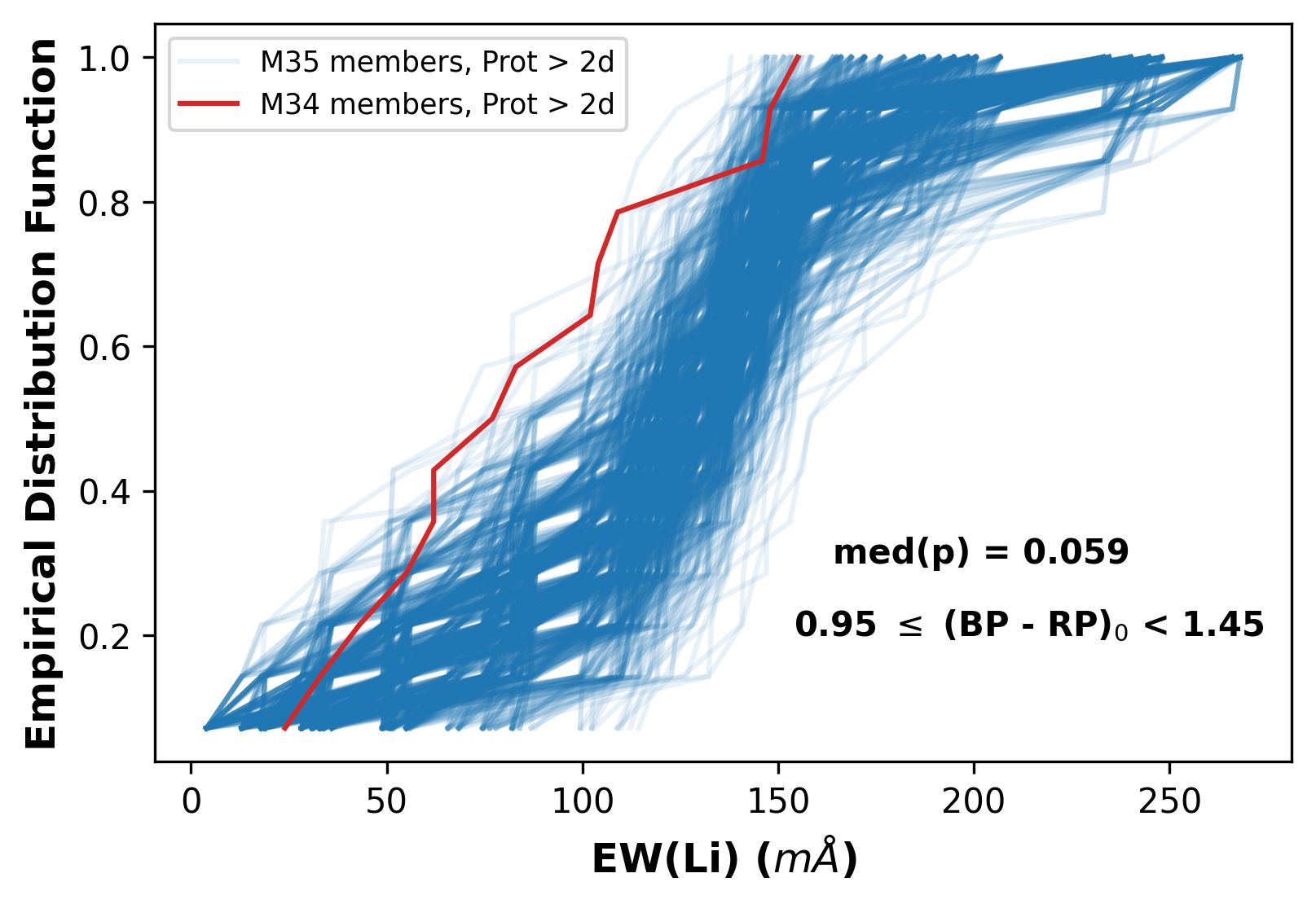}
   \caption{Empirical distribution functions obtained from 1000 Kolmogorov-Smirnov tests performed between the Li EWs sample corresponding to M35 and those of other clusters. \textit{Left---} Pleiades Li EWs sample (see Figures~\ref{EWLi_Teff_M35_Pleiades} and \ref{Prot_Teff_M35_Pleiades}). For each test we have randomly picked 48 Li EWs from the M35 sample (blue), that is, the size of the Pleiades sample, and compared them with the Pleiades sample (black). \textit{Right---} M34 Li EWs sample (see Figure \ref{Prot_Teff_M35_M34}). For each test we have randomly picked 14 Li EWs from the M35 sample (blue), that is, the size of the M34 sample, and compared them with the latter (red). The range of \textit{Gaia} colours employed in each case is written in each panel. We selected G and K stars for the comparison between M35 and the Pleiades, but only K stars in the comparison with M34 due to the few data available for this cluster.}
              \label{test_KS}%
    \end{figure*}

To statistically compare the Li measurements collected for the three clusters, we have performed Kolmogorov-Smirnov tests between them \citep{Hodges1958}. The purpose of this test is to verify whether two samples (in this case, the Li EWs collected) have been taken from the same continuous distribution. We have only considered single stars with rotation periods $>$2 d (to avoid contamination from rapid rotators) and absolute Li EWs (not upper values). Figure \ref{test_KS} shows the empirical distribution functions obtained for each of the 1000 tests performed between the Li EWs sample corresponding to M35 and the samples corresponding to the Pleiades and M34. The median of the p-values obtained in both cases (shown in the panels of the figure) do not allow us to rule out the possibility that the three samples come from the same continuous distribution, implying a clear similarity between them. On the other hand, the median of the p-values obtained when we compared the Pleiades sample and the M34 sample in the same way is higher, indicating a larger difference between the Li distributions of these two clusters. 

\section{Summary and conclusions}\label{sec:conclusions}

We collected a sample of 396 G and K dwarfs observed with the Hydra spectrograph of the 3.5-m telescope located at WYIN Observatory. We analysed the photometry, proper motions, and radial velocities of these sources, identifying 251 members of the M35 open cluster, 209 of which are single stars. In addition, we obtained rotation periods for 47 of these sources from ZTF light curves and we took advantage of previous photometric surveys, obtaining rotation periods for 197 M35 members in total. 

We analysed the spectra collected for 110 out of 251 M35 members, deriving iron abundances, Li EWs and non-LTE Li abundances for them. From the iron abundances obtained for the most probable single members with effective temperatures higher than 4500 K we derived a metallicity of [Fe/H]~=~$-$0.26$\pm$0.09 for M35. In addition, we found that the empirical law proposed in \cite{soderblom1993} for the Pleiades overestimates the equivalent width of the Fe I line at 670.75 nm in 5 - 15 m$\AA$ for the range of effective temperatures analysed. As a result, we conclude that the \cite{soderblom1993} empirical relation is not recommended for subsolar metallicity clusters. 

We combined the Li EWs that we measured with data collected from previous studies, obtaining rotation periods and Li EWs for 257 single members of M35. The Li--rotation connection found for this sample follows the trend observed in similar studies of M35 as well as with the trend found for other stellar associations: fast rotators are richer in Li than slower rotators of the same effective temperature.

Finally, we compared our sample with that for two open clusters with solar metallicity: the Pleiades, which is slightly younger than M35, and M34, which is slightly older. From the Kolmogorov-Smirnov tests performed, no significant statistical difference was observed between the Li EWs distribution for M35 and those for the other clusters. However, a larger difference is found between the M34 sample and the Pleiades one. As a result, we conclude that a 0.2 - 0.3~dex difference in metallicity does not have an observable impact on the Li distributions of open clusters with ages between 100 and 250~Myr. The statistical analysis carried out points to age as the parameter that most influences the evolution of Li in open clusters.


\begin{acknowledgements}
      This research has been funded by grant No. PID2019-107061GB-C61 by the Spanish Ministry of Science and Innovation/State Agency of Research MCIN/AEI/10.13039/501100011033 and No. MDM-2017-0737 Unidad de Excelencia “María de Maeztu”- Centro de Astrobiología (INTA-CSIC). Based on Clusterix 2.0 service at CAB (INTA-CSIC). This publication makes use of VOSA, maintained under the Spanish Virtual Observatory project funded by MCIN/AEI/10.13039/501100011033/ through grant PID2020-112949GB-I00. VOSA has been partially updated by using funding from the European Union's Horizon 2020 Research and Innovation Programme, under Grant Agreement nº 776403 (EXOPLANETS-A). 

      D.C.M. acknowledges support from a Erasmus+ grant co-funded by the European Commission and the Spanish Ministry of Education, Culture and Sport. D.C.M. thanks Dr L. Casamiquela, Dr M. Cerviño, Dr E. Marfil, Dr M. Morales-Calderón and A. Saavedra (CAB) for their comments and suggestions.

      M.A.A.~acknowledges support from a Fulbright U.S.~Scholar grant co-funded by the Nouvelle-Aquitaine Regional Council and the Franco-American Fulbright Commission. M.A.A.~also acknowledges support from a Chr\'etien International Research Grant from the American Astronomical Society.

      Based on observations obtained with the Samuel Oschin Telescope 48-inch and the 60-inch Telescope at the Palomar Observatory as part of the Zwicky Transient Facility project. ZTF is supported by the National Science Foundation under Grants No. AST-1440341 and AST-2034437 and a collaboration including current partners Caltech, IPAC, the Weizmann Institute for
Science, the Oskar Klein Center at Stockholm University, the University of Maryland, Deutsches Elektronen-Synchrotron and
Humboldt University, the TANGO Consortium of Taiwan, the University of Wisconsin at Milwaukee, Trinity College Dublin,
Lawrence Livermore National Laboratories, IN2P3, University of Warwick, Ruhr University Bochum, Northwestern University and
former partners the University of Washington, Los Alamos National Laboratories, and Lawrence Berkeley National Laboratories.
Operations are conducted by COO, IPAC, and UW.

\end{acknowledgements}


%
\bibliographystyle{aa} 
\bibliography{M35_Lithium_v5} 
%

\end{document}